\documentclass[fleqn,usenatbib]{mnras}

\usepackage{newtxtext,newtxmath}

\usepackage[T1]{fontenc}

\DeclareRobustCommand{\VAN}[3]{#2}
\let\VANthebibliography\thebibliography
\def\thebibliography{\DeclareRobustCommand{\VAN}[3]{##3}\VANthebibliography}

\usepackage{graphicx}	
\usepackage{amsmath}	
\usepackage{xcolor}     

\newcommand{\angstrom}{\mbox{\normalfont\AA}}

\title[3D NLTE Supernova Spectra]{Modelling supernova nebular lines in 3D with \texttt{ExTraSS}}

\author[B.F.A. van Baal et al.]{
Bart F. A. van Baal,$^{1}$\thanks{E-mail: barteld.vbaal@astro.su.se}
Anders Jerkstrand,$^{1}$
Annop Wongwathanarat$^{2}$
and Hans-Thomas Janka$^{2}$
\\
$^{1}$The Oskar Klein Centre, Department of Astronomy, Stockholm University, AlbaNova, Se-10691 Stockholm, Sweden\\
$^{2}$Max Planck Institute for Astrophysics, Karl-Schwarzschild-Str 1, D-85748 Garching, Germany\\
}

\date{Accepted XXX. Received YYY; in original form ZZZ}

\pubyear{2023}

\begin{document}
\label{firstpage}
\pagerange{\pageref{firstpage}--\pageref{lastpage}}
\maketitle

\begin{abstract}  
We present \texttt{ExTraSS} (EXplosive TRAnsient Spectral Simulator), a newly developed code aimed at generating 3D spectra for supernovae in the nebular phase by using modern multi-dimensional explosion models as input. It is well established that supernovae are asymmetric by nature, and that the morphology is encoded in the line profiles during the nebular phase, months after the explosion. In this work, we use \texttt{ExTraSS} to study one such simulation of a $3.3\,M_\odot$ He-core explosion ($M_\text{ejecta}=1.3\,M_\odot$, $E_\text{kin}=1.05\times10^{51}\,$erg) modelled with the \texttt{Prometheus-HotB} code and evolved to the homologous phase.
Our code calculates the energy deposition from the radioactive decay of $^{56}$Ni $\rightarrow$ $^{56}$Co $\rightarrow$ $^{56}$Fe and uses this to determine the Non-Local-Thermodynamic-Equilibrium temperature, excitation and ionization structure across the nebula. From the physical condition solutions we generate the emissivities to construct spectra depending on viewing angles. Our results show large variations in the line profiles with viewing angles, as diagnosed by the first three moments of the line profiles; shifts, widths, and skewness. We compare line profiles from different elements, and study the morphology of line-of-sight slices that determine the flux at each part of a line profile. We find that excitation conditions can sometimes make the momentum vector of the ejecta emitting in the excited states significantly different from that of the bulk of the ejecta of the respective element, thus giving blueshifted lines for bulk receding material, and vice versa. We compare the 3.3 $M_\odot$ He-core model to observations of the Type Ib supernova SN 2007Y. 
\end{abstract}




\section{Introduction}
Core-collapse supernovae (CCSNe) are the final step in the evolution of massive stars \citep[$M_{\rm ZAMS} \gtrsim 8\,$M$_\odot$,][]{heger2003massive}. 
These explosions leave behind a central compact object (either a black hole or a neutron star), while expelling the rest of the star, enriching the cosmos in elements produced both during hydrostatic and explosive burning \citep{arnett1996supernovae,woosley2002evolution}. The expanding debris is radioactively powered and transitions from an initial optically thick diffusion phase to a later more optically thin ("nebular") phase in which emission lines can be seen and decoded to infer the structure of the supernova. 

Multiple lines of observational evidence indicate that SNe are strongly asymmetric by nature.
Nearby spatially resolved supernova remnants show complex structures with plumes, knots and filaments \citep[e.g.][]{hughes2000nucleosynthesis,fesen2006expansion,larsson2013morphology}. Emission line profiles often show deviations from symmetry \citep{filippenko1989spectroscopic}. Observed high kick velocities of neutron stars also indicate asymmetric explosion dynamics \citep{arzoumanian2002velocity,Hobbs2005statistical}, as do results from spectropolarimetry \citep{Tanaka2012,Reilly2016,Tinyanont2021}.

The explosion itself may give energy deposition and shock expansion stronger in certain directions than others. In the case of neutrino-driven explosions, for example, the neutrino energy deposition triggers Rayleigh-Taylor instability in the neutrino-heated postshock layer and the initially stalled supernova shock is also subject to nonradial deformation modes \citep[see e.g.][]{herant1994postexplosion,burrows1995nature,janka1996neutrino,blondin2003stability,buras2006two,blondin2007pulsar,takiwaki2014comparison,lentz2015three,melson2015neutrinoiron,melson2015neutrino20modot,janka2016physics,roberts2016general,wongwathanarat2017production,ott2018progenitor,vartanyan2019successful,muller2020hydrodynamics}. In addition to this, Rayleigh-Taylor unstable conditions are created as the shock passes certain shell interfaces on its way out \citep{chevalier1976hydrodynamics,shigeyama1990low,muller1991instability,hachisu1991rayleigh,nomoto1995evolution,kifonidis2006nonspherical}. The growth of these secondary Rayleigh-Taylor instabilities is seeded and boosted by the initial explosion asymmetries \citep[as discussed in detail in][]{kifonidis2003nonspherical,kifonidis2006nonspherical,hammer2010three,wongwathanarat2015three}. These two effects combine to produce complex 3D structures. Interaction with an asymmetric stellar environment (e.g. collision with a disk/CSM or a binary companion), may also introduce further asymmetries in some supernovae.

With steady improvements in the computational modelling of both the explosion and subsequent instabilities, there are now realistic 3D models available which model the ejecta for $\gtrsim1\,$day \citep[e.g.][]{wongwathanarat2015three,wongwathanarat2017production,muller2018multidimensional,stockinger2020three,gabler2021infancy}, and efforts to produce observables from these to compare to observations are underway \citep{Alp2019,muller20203D,jerkstrand2020properties,fields20213D,gabler2021infancy,orlando2021fully,kozyreva2022low}. 

A focus on inferring the multi-dimensional structure of SNe from nebular-phase line profiles began in earnest with the Type Ic-BL SN 1998bw. Its association with GRB 980425 made deviation from spherical symmetry relevant from the start, and its emission lines could only be understood by models taking this into account \citep{mazzali2001nebular, maeda2003bipolar, maeda2006optical}.

\citet{maeda2008asphericity} and \citet{modjaz2008double} showed that significant line asymmetries are present also in many regular Type Ibc SNe, not just those associated with GRBs.
\citet{taubenberger2009nebular} investigated a large sample and developed a taxonomy for different line profile types. In these studies a large variety of [O I] line profiles was uncovered. \citet{maeda2008asphericity} found a high incidence of double-peaked profiles. \citet{taubenberger2009nebular} found that at very late times the profile of the Mg I] emission generally resembles the [O I] profile when accounting for the doublet nature of the latter, thereby giving constraints on the relative 3D distributions of O and Mg. 

A recent study by \citet{fang2022core} looked at correlations between [O I] and [Ca II] profiles in order to connect the progenitor CO core masses with the ejecta geometry. There is now enough data, and available 3D hydrodynamic models evolved to late times, to tackle the interpretation of nebular line profiles at depth. This requires the development of realistic models for the nebular-phase emission in 3D.

Spectral modelling in the nebular phase is a complex task even in 1D. The reason is that the ejecta are in a regime where densities are low enough that Non-Local Thermodynamic Equilibrium (NLTE) is needed, while not low enough that radiative transfer effects can generally be ignored as e.g. photoionization can play an important role, and some lines and wavelength regions can remain optically thick for several years \citep{jerkstrand201144ti,jerkstrand2011PhDspectral}. In addition to this, the energy cascade spans over 6 orders of magnitude, from injection of radioactive decay particles at MeV energies down to cooling in the optical and infrared at $\lesssim$ 1 eV energies. Modelling therefore requires large computational efforts typically involving the iterative solution of non-thermal electron degradation, rate equations for $\gtrsim 10^5$ levels, and radiative transfer in $\gtrsim 10^6$ lines and continua.

One such modelling framework is the \texttt{SUMO} code \citep[SUpernova MOnte Carlo,][]{jerkstrand201144ti,jerkstrand2012progenitor}, which is a NLTE spectral synthesis code which employs both detailed microphysics and line-by-line radiative transfer to create spectral outputs. Recently, \texttt{SUMO} was upgraded to be able to handle also molecular effects \citep{liljegren2020carbon,liljegren2022molecular}, magnetar powering \citep{omand2022towards}, and r-process physics for kilonova applications\footnote{A kilonova is the optical counterpart of two merging neutron stars.} \citep{pognan2022validity,pognan2022NLTE}.

\texttt{SUMO} uses its Monte Carlo foundation to allow for an innovative statistical description of mixing effects, critically by allowing chemically distinct clumps to habitate the same radial velocity ranges, as seen in multi-D hydro simulations. This feature is critical to achieve realistic supernova spectra. It is, nevertheless, a fundamentally 1D code $-$ and with the rise of multi-dimensional explosion models \citep[see e.g.][for recent reviews]{muller2016status,janka2016physics,janka2017neutrino,muller2020hydrodynamics,burrows2021corecollapse} the scientific value of computing spectral predictions directly from these is apparent.

\citet{jerkstrand2020properties} made the first step to consider multiple dimensions by developing a new platform capable of reading in 3D hydrodynamic models, parameterizing the emissivity in each cell, and performing both Monte Carlo and ray-tracing radiative transfer with a simple opacity. One of the important code features developed was the ability to transfer photon packets on a spherical coordinate grid instead of a Cartesian grid as traditionally used in 3D codes. \citet{jerkstrand2020properties} used the code to study the properties of gamma-ray deposition and emergent gamma-ray decay lines for Type II supernova models in 3D, which can be treated with the physics implemented as of 2020.

In this work, we build further on these first steps and develop the 3D platform to be able to compute the NLTE physical conditions and full UVOIR emissivity in each cell. We give the official name \texttt{ExTraSS} (EXplosive TRAnsient Spectral Simulator) to this new code. In this paper we present the physics added to the code, and apply it to a single 3D Type Ib supernova model. We present the first viewing-angle distributions of fundamental nebular line metrics - shift, width, and skewness - as obtained from a current state-of-the-art core-collapse explosion model in 3D, and discuss how these compare to observations.

The paper is organized as follows. In Section \ref{sec:Methods_code} we cover in detail the setup of \texttt{ExTraSS}, its new features, and the relevant physics. In Section \ref{sec:Methods_explosion} we outline the Type Ib explosion model and setup. In Section \ref{sec:Results} we validate outputs of our new code and showcase the first results. In Section \ref{sec:ObsComp} we compare our synthetic spectra against observational spectra of SN 2007Y, which has similar ejecta properties as our model. In Section \ref{sec:Discussion} we evaluate our findings and highlight what comes next. In Section \ref{sec:Conclusion} we recap the main findings of our work.

\section{ \texttt{ExTraSS} - a spectral synthesis code in 3D} \label{sec:Methods_code}
The starting point of this work was the 3D platform developed by \citet{jerkstrand2020properties} which  considers internal energy deposition and emergent $\gamma$-ray lines. We here significantly expand this code to be able to compute temperature and NLTE level populations, including non-thermal physics.

A major challenge is the sheer size of the problem. In 1D codes a setup might entail $\mathcal{O}(10^2)$ zones, while our 3D input models generally have as many cells as $\mathcal{O}(10^{7-8})$. 
Even if we reduce each dimension of the input models by a factor of 3 or 4 by downsampling, we still end up with $\mathcal{O}(10^{5-6})$ cells. 
As a result, care has to be taken in the design both for RAM and run-time considerations.

To keep the size of the problem limited, we currently restrict ourselves to using a maximum of $100$ excitation states per ion (but some have fewer). The most important states for the nebular phase are typically the low-lying ones, so we expect this limitation not to have a great impact on accuracy. For ionization we currently allow for the first three ionization states for each element; this means that up to 300 excitation states are considered per element, if each ion has 100 levels. 

The first step in \texttt{ExTraSS} is the gamma deposition; this is fully described in \citet{jerkstrand2020properties}. 
Once this process has been completed, 
the calculated energy deposition is used combined with the composition in each cell to find the Non-Local Thermodynamic Equilibrium (NLTE), which gives the level populations and temperature in each cell. In order to calculate the excitation and ionization structures of each element, we have replaced the separate excitation and ionization solvers in \texttt{SUMO} with a new solver ('excion$\_$solver') coupling these together. This enables improved convergence properties and fewer iterations. This new excitation-ionization solver is then iterated with the standard temperature solver and Spencer-Fano subroutine \citep{spencer1954energy} of \texttt{SUMO}.

In Table \ref{tab:NLTE_table} a schematic overview is given for a hypothetical atom with two ionization stages, each with three excitation states. The table shows the full excitation-ionization matrix which describes the transitions between the states, together with the level population vector. The first rate equation is replaced with the number conservation equation for the element to close the system. In \texttt{ExTraSS} this matrix is slightly simplified\footnote{For most ions the distribution of ionization target states are not known; but e.g. for O I $\rightarrow$ O II data shows that about 2/3 of ionizations go the GS at relevant energies \citep{laher1990updated}. For recombinations, for nebular SNe most recombining ions will be in the ground state.}; ionization flows are always allocated to the ground state of the ion, so $\Gamma_{ij}=0$ for $j\neq\text{ion}_{\text{gs}}$. Additionally, our recombination rates R$_{ij}$ only depend on $j$ (state in the recombined ion) and are the same for all starting states $i$.

In this paper we include non-thermal and thermal collisional processes (excitation and ionization), spontaneous radiative decay (with Sobolev escape probabilities), and recombination. We do not include photoexcitation, deexcitation or photoionization by the diffuse radiation field. The atomic data used is the same as in \texttt{SUMO}. If the atomic data is missing, the approximate treatments from \texttt{SUMO} are used.
\begin{table}
	\centering
     \setlength\tabcolsep{3pt}  
     \begin{tabular}{cccc}
        $\left[ \begin{array}{ccc ccc} 
        1 & 1 & 1 & 1 & 1 & 1 \\
	 	\text{X}_{12} &\text{-T}_{22} & \text{X}_{32} & \text{R}_{42} & \text{R}_{52} & \text{R}_{62} \\
	 	\text{X}_{13} & \text{X}_{23} &\text{-T}_{33} & \text{R}_{43} & \text{R}_{53} & \text{R}_{63} \\
	 	\Gamma_{14} & \Gamma_{24} & \Gamma_{34} &\text{-T}_{44} & \text{X}_{54} & \text{X}_{64}\\
	 	\Gamma_{15} & \Gamma_{25} & \Gamma_{35} & \text{X}_{45} &\text{-T}_{55} & \text{X}_{65}\\
	 	\Gamma_{16} & \Gamma_{26} & \Gamma_{36} & \text{X}_{46} & \text{X}_{56} &\text{-T}_{66}\\ \end{array}\right]$ 
            &
        $\left[ \begin{array}{c} n_1 \\ n_2 \\ n_3 \\ n_4 \\ n_5 \\ n_6 \end{array}\right]$ & = & $\left[ \begin{array}{c} n_\text{element} \\ 0 \\ 0 \\ 0 \\ 0 \\ 0 \end{array}\right]$
     \end{tabular}

	\caption{A schematic overview of what the excitation-ionization matrix calculation looks like for an atom with two ionization stages (neutral corresponding to levels $1-3$ and ionized to levels $4-6$). The subscripts $_{ij}$ give the flow from  state $i$ to state $j$. The diagonal values T$_{ii}$ contain the sum of outgoing flows from $i$, hence the minus sign. X refers to excitation/deexcitation flows, $\Gamma$ to ionization flows (from neutral to singly ionized in this example) and R to recombination flows (from singly ionized to neutral in this example). $n_i$ is the vector of current level populations for which the matrix is solved, while on the right-hand side the conservation equation is given together with the first row on the left-hand side.}
	\label{tab:NLTE_table}
\end{table}

The excitation/ionization solver solves for every element\footnote{There are 14 elements in our explosion model: He, C, O, Ne, Mg, Si, S, Ar, Ca, Ti, Cr, Fe and Ni. Additionally, an element "X" is considered as catch-all element for iron-group elements of unspecified detail; for our solver element X is combined with Ni. This leaves us with (at most) 13 elements in any cell.} independently in every iteration, for up to 400 iterations\footnote{Convergence failure can occur but are rare. Non-converged cells are flagged nad removed from the spectrum generation step. We check that the non-converged cells have a negligible fraction of the total $\gamma$ deposition.}. For each element, in each iteration, the excitation-ionization matrix as in Table \ref{tab:NLTE_table} is constructed and then a numerical Jacobian is calculated. The non-linearity comes from the Sobolev escape probabilities, which depend on the lower level population. The Jacobian is used to solve for the next step, using a dampening factor, in a Newton-Raphson scheme. After each element has obtained a new excitation and ionization structure, the free electron fraction is updated such that the next element has the most recent value. Every element is checked for convergence separately, and only non-converged elements are still looped over in each excion-iteration. The convergence criterium for an individual element is that the new level populations are less than $1\%$ different from the old ones for every state. The Spencer-Fano subroutine, which determines what fraction of the energy goes into ionizing each element, is called on the first two iterations, and then every fourth iteration. This is for computational reasons as it is a time-consuming step, but tends to change little past the first two iterations. 

The dampening factor used in \texttt{ExTraSS} is not constant; for the first 40 iterations it is 0.8, then for the next 80 iterations it is 0.16 and in the remainder of the iterations it is 0.025. We chose this scheme as some elements have a hard time converging when the dampening factor is large, but few elements make it past the first 40 iterations which means only for a few outliers the non-default value is used. After every element is marked as converged, the temperature solver is invoked to evaluate the heating and cooling using the new excitation and ionization structures calculated. The temperature is then modified before looping back into the excion$\_$solver where all elements are flagged as not-yet converged, and another excion-loop begins. Once the excion$\_$solver and the temperature solver both find convergence, the cell as a whole is marked as converged, which means it has found its NLTE solution.

A failsave is also built into the code for cells which have very little energy deposition which can cause numerical issues for the level population solver. Generally, these cells also have very low mass so their exclusion should not make a large impact on the final spectrum. The failsave is checking specifically that a cell retains a high enough temperature and free electron fraction ($x_e$); any cell which drops below $T<25\,$K is flagged, as well as cells which drop below $T<500\,$K and $x_e<0.03$. These thresholds are chosen as typically the cells with too little energy deposition cannot balance both temperature and ionization stages so they both spiral down completely or occasionally the temperature tries to hit absolute 0. 

Upon convergence of all cells, \texttt{ExTraSS} takes the level populations and calculates the emissivity from each cell. Once this is completed, for each viewing angle the Doppler-shift corrected flux at the corresponing Doppler-shifted wavelength is added to the correct wavelength bin. For the current version of \texttt{ExTraSS}, radiative transfer is not considered (beyond the local Sobolev escape probabilities used in the rate equations) and thus we assume that all emission which is sent out from a cell will reach the observers.

The way in which \texttt{ExTraSS} determines the emissivity (at rest wavelength $\lambda$, initially) from each cell is by using the following formula: 
\begin{equation}
    \text{Emissivity}(\lambda) = A_{ul} * \Delta E_{ul} * \beta_{S,ul} * N_u,
    \label{eq:emis}
\end{equation}
where $A_{ul}$ refers to the transmission strength from the upper excitation state $_u$ to the lower state $_l$, $\Delta E_{ul}$ to the energy difference between the two states, $\beta_{S,ul}$ to the Sobolev escape probability (set by the Sobolev optical depth) from this upper level to the lower level, and $N_u$ to the total amount of particles in this excitation state in this cell. This emissivity value is stored alongside the rest wavelength of this transmission, creating arrays which are different for each cell. 

With the emissivities in each cell determined, the final spectrum for each viewing angle can be obtained. This is done by calculating the Doppler shifts for each cell with regards to the viewing angle, and then binning the Doppler shifted emissivity into the correct, Doppler shifted wavelength bin for each viewing angle, generating separate spectra for each observer. To create a smoother Doppler shift profile, each cell is split into 8 sub-parts which send out 1/8th of the emissivity, which numerically smooths the velocity distribution in the grid and thus also smooths the spectra. A test with 27 sub-parts showed no further improvement in smoothing but did have a negative impact on the runtime, and thus the 8-point splitting was adopted.

\section{Explosion and hydrodynamic modelling} \label{sec:Methods_explosion}

\subsection{Explosion Code} \label{ssec:explosioncode}
The simulations of the explosion are done with the finite-volume Eulerian multifluid hydrodynamics code \texttt{Prometheus} \citep{fryxell1991instabilities,muller1991instability,muller1991high}. The multidimensional Euler equations are integrated with the dimensional splitting technique of \citet{strang1968construction}, the code uses the piece-wise parabolic method (PPM) of \citet{colella1984piecewise} together with the exact, iterative Riemann solver for real gases of \citep{colella1985efficient}. For grid cells where strong grid-aligned shocks are present, the AUSM+ Riemann solver of \citet{liou1996sequel} is used, as this prevents numerical artifacts created through the odd-even decoupling \citep[see][]{quirk1994contribution}. The specific version used is the \texttt{Prometheus-HotB} version \citep{janka1996neutrino,kifonidis2003nonspherical,kifonidis2006nonspherical,scheck2006multidimensional,arcones2007nucleosynthesis,wongwathanarat20133D,wongwathanarat2015three,wongwathanarat2017production,gessner2018hydrodynamical,stockinger2020three}, which includes neutrino physics, a general equation of state that is applicable both above and below the nuclear statistical equilibrium, and a small $\alpha$ network to treat nuclear burning. The code uses an approximate, gray treatment for neutrino transport \citep{scheck2006multidimensional} which is connected to the combined luminosity of all species at a contracting Lagrangian radius at $1.1\,M_\odot$, so firmly inside the neutrinosphere. This allows for the tweaking of the explosion energy to (close to) the desired value \citep[see also][]{wongwathanarat2017production}. The neutrino mechanism, combined with the hydrodynamic instabilities, is responsible for creating the pronounced asymmetries in the first second of the explosion and leaves its imprints behind as large-scale asphericities on the shock wave, innermost ejecta and overall explosion. 

The spatial discretization utilizes the Yin-Yang overlapping grid technique in spherical geometry \citep{kageyama2004yin,wongwathanarat2010axis}. This technique avoids numerical artifacts that can arise near the polar axis of a spherical polar grid, and it also alleviates time-step constraints imposed by the Courant-Friedrich-Levy (CFL) condition. The CFL condition in a spherical polar grid is very limiting due to the small azimuthal grid cells at the polar regions, but with the Yin-Yang setup this condition does not apply and larger time-steps can be utilized. 

The $\alpha$ network used contains 14 species: the $\alpha$ nuclei from $^4$He to $^{56}$Ni and an additional 'Element X' which is used to track the production of neutron-rich nuclear species at lower electron fractions ($Y_e < 0.49$). Element X is effectively a catch-all for iron-group elements of unspecified detail, for which it becomes very expensive to exactly determine the outcome of the nuclear burning.

\subsection{Ejecta modelling} \label{ssec:ejectamodel}
\begin{figure}
    \centering
    \includegraphics[width=\linewidth]{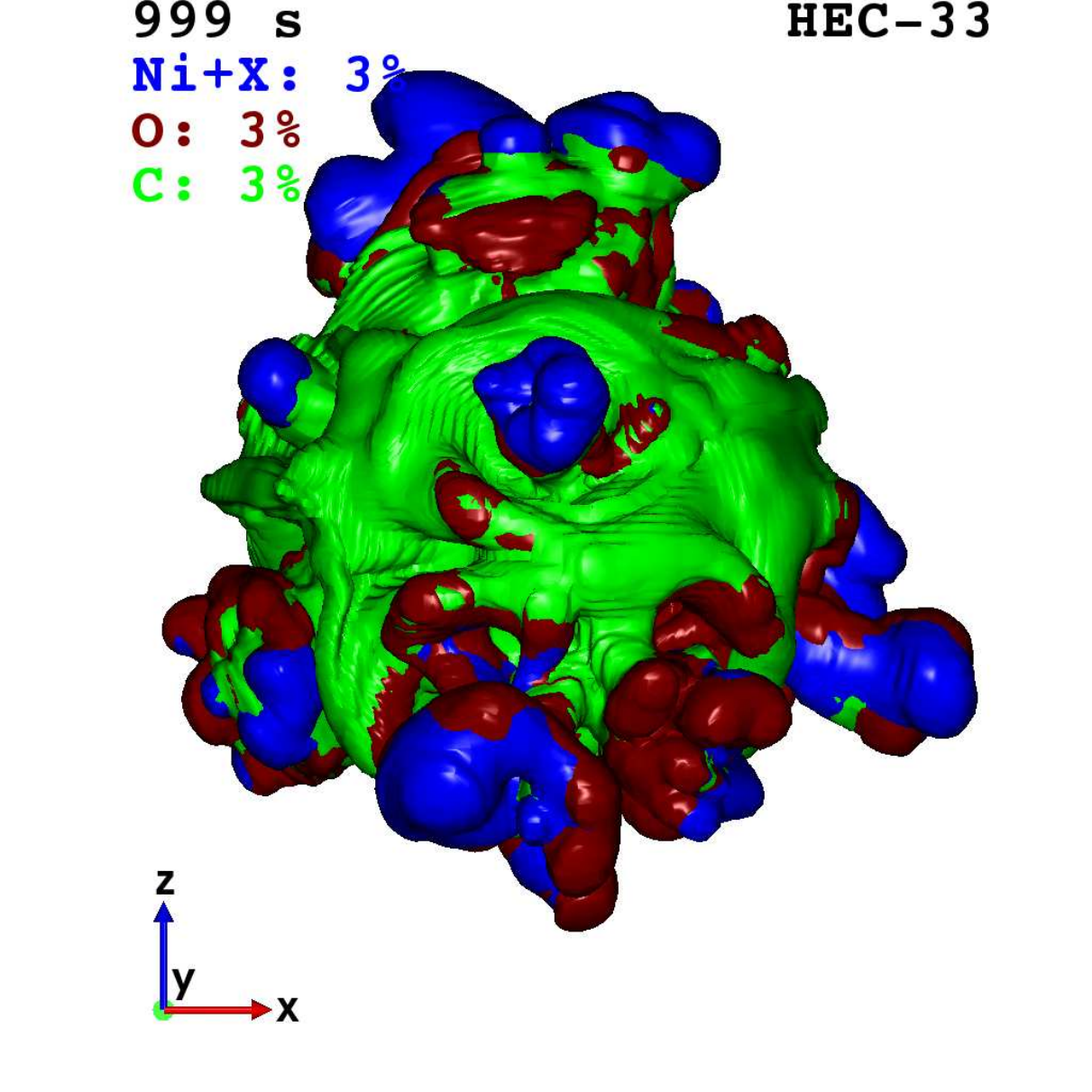}
    \caption{A 3D rendering of the explosion model, 'HEC-33', at 999s post explosion. The model is a $3.3\,M_\odot$ He-core star exploded with 1.05 B. The isosurfaces of $3\%$ mass fraction of nickel + element X (blue), oxygen (red) and carbon (green) are shown. The xyz-tripod is shown in the bottom left corner, with the positive y-axis directed into the figure.}
    \label{fig:exploModel3D}
\end{figure}
\begin{figure*}
    \centering
    \includegraphics[width=\linewidth]{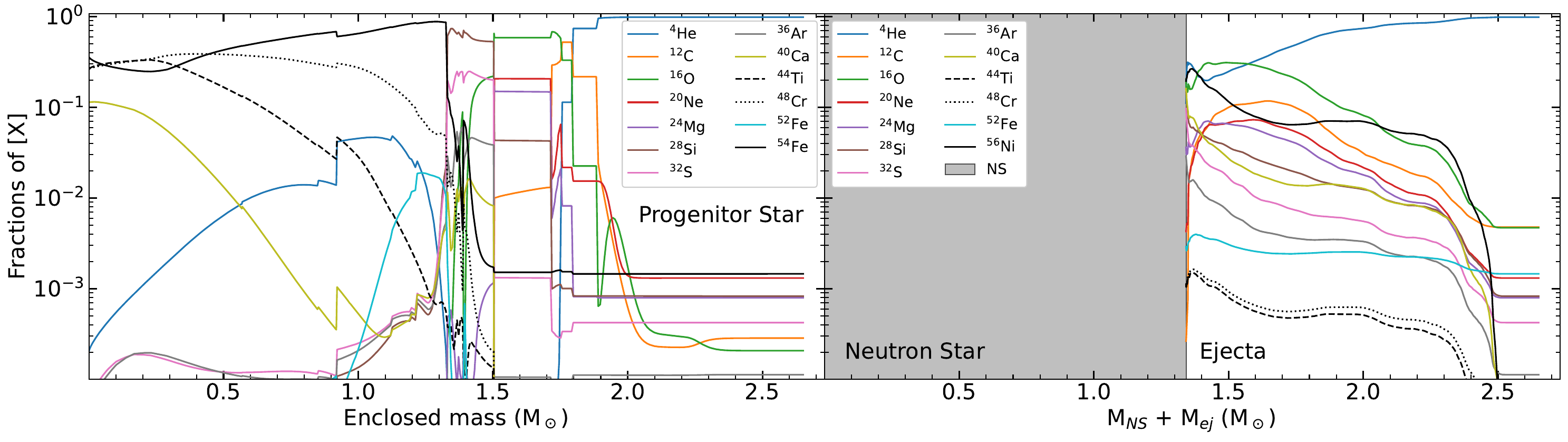}
    \caption{A comparison between the chemical compositions of the progenitor star (left) and ejecta (right) against their mass coordinates. On both sides the 12 $\alpha$ nuclei from $^{4}$He to $^{52}$Fe are shown, together with $^{54}$Fe for the progenitor (the dominant component of the Fe-core) and $^{56}$Ni for the ejecta (which is the source of $\gamma$-rays through radioactive decay). The shaded area on the right side indicates the mass which has formed the NS ($M_\text{NS}=1.34\,M_\odot$).}
    \label{fig:comp_HEC-33_compare}
\end{figure*}
The ejecta are tracked on a spatially fixed grid of size $N_r\times N_\theta\times N_\phi$, where $N_\theta=90$ and $N_\phi=180$ leading to a uniform angular resolution of $2^{\circ}$ for all angular directions (and implying a number of $47 \times 137$ angular grid cells for each of the two Yin-Yang subgrids), just as in \citet{jerkstrand2020properties}. The amount of radial zones $N_r$ is flexible; the outer boundary is always set to $r_\text{out}=2\times10^{10}\,$km but the inner boundary moves forward in time during the simulation; mass leaving the grid through the inner boundary is assumed to be accreting back onto the central object as fallback. 

For the purpose of testing and demonstrating \texttt{ExTraSS}, we have chosen to study one model in detail; in a forthcoming paper we will analyse many more. The model for this study is a $3.3\,M_\odot$ helium star progenitor, based on \citet{ertl2020explosion}'s $3.3\,M_\odot$ He-star, which was evolved with a standard mass loss recipe to the presupernova stage by \citet{woosley2019evolution}. To break the symmetry in the original 1D model, random perturbations were imposed on the radial velocity, similar to \citet{wongwathanarat2010hydrodynamical,wongwathanarat20133D}. In the rest of this work, we will refer to the model as HEC-33.

HEC-33 has evolved to a pre-SN mass of $M_\text{pre-SN}=2.67\,M_\odot$ with a CO-core mass of $M_\text{CO}=1.75\,M_\odot$ and an iron core of $M_\text{Fe}=1.33\,M_\odot$ by the time of explosion. The star explodes with $E_\text{kin}=1.05\,$B (B$=10^{51}\,$erg), creating a neutron star with $M_\text{NS}=1.34\,M_\odot$ and an ejecta with a mass of $M_{ej}=1.3\,M_\odot$. About $2/3$rds of the ejecta is He, and there is $0.084\,M_\odot$ Ni+X in the ejecta. At the end of the \texttt{Prometheus-HotB} simulation, $N_r=2512$, the ejecta has been modelled for $999\,$s since the initial explosion. A 3D rendering of the ejecta at this time is shown in Figure \ref{fig:exploModel3D}. At the end of the explosion the neutron star has achieved a Cartesian velocity of $v_x=62\,\text{km}\,\text{s}^{-1}$, $v_y=-4.7\,\text{km}\,\text{s}^{-1}$, $v_z=-120\,\text{km}\,\text{s}^{-1}$, for a total NS kick of $135\,\text{km}\,\text{s}^{-1}$.

The level of mixing induced by the explosion compared to the structure of the progenitor is quite drastic. In Figure \ref{fig:comp_HEC-33_compare} the compositions of the progenitor star and the ejecta (at $t=999\,$s) are shown side by side to highlight this degree of mixing. It can clearly be seen that part of the $^{56}$Ni has achieved high velocities as it is present throughout most of the ejecta, except for the outer $\sim0.2\,M_\odot$ of the ejecta which has retained a pure He-composition. Almost all the other elements have some presence in the first $\sim1\,M_\odot$ of the ejecta, although contributions from $^{44}$Ti, $^{48}$Cr and $^{52}$Fe are minimal. We give more details on the progenitor structure in Appendix \ref{app:progenitor}.

One of the critical points is the mapping from the \texttt{Prometheus-HotB} code to \texttt{ExTraSS}, as this is the point where we no longer model the hydrodynamical behaviour of the ejecta and instead assume we can homologously evolve the ejecta with a purely radial velocity component into the nebular phase\footnote{In the nebular phase this behaviour is also often called 'homologeous expansion', and although our explosion model might not have fully reached the true free-coasting limit yet we will make the assumption that this expansion can be applied.}. We asses the degree of homology as well as the fallback mass at $t=999\,$s in Appendix \ref{app:homology_confirm}, to validate that our fast-forwarding is not done too early and the model converges.

\section{Results} \label{sec:Results}
In this section we are going to review the first results of our new code. As the ejecta are homologous, the natural independent variable to plot quantities against is velocity.

We will investigate the properties of three main metrics of an emission line; \emph{line centroid shift} ($v_\text{shift}$), \emph{line width} ($v_\text{width}$), and \emph{skewness} ($v_\text{skew}$). These correspond to the first, second, and third moments of the line profile and are calculated as follows:
\begin{equation}
    v_\text{shift} = \dfrac{\int_{-v_{max}}^{v_{max}} L_\lambda v(\lambda) d\lambda}{\int_{-v_{max}}^{v_{max}} L_\lambda d\lambda},
    \label{eq:centroidEQ}
\end{equation}
\begin{equation}
    v_\text{width} = 2.35 \times \Bigg( \dfrac{\int_{-v_{max}}^{v_{max}} [v(\lambda) - v_\text{shift}]^2 L_\lambda d\lambda}{\int_{-v_{max}}^{v_{max}} L_\lambda d\lambda} \Bigg)^{1/2},
    \label{eq:widthEQ}
\end{equation}
\begin{equation}
    v_\text{skew} = \Bigg( \dfrac{\int_{-v_{max}}^{v_{max}} [v(\lambda) - v_\text{shift}]^3 L_\lambda d\lambda}{\int_{-v_{max}}^{v_{max}} L_\lambda d\lambda} \Bigg)^{1/3}.
    \label{eq:skewEQ}
\end{equation}
In each of these equations, $L_\lambda$ is the spectral luminosity, and $v(\lambda)$ is given by the offset of the observed wavelength ($\lambda$) to the rest wavelength ($\lambda_0$): $v(\lambda) = c \times \left(\lambda - \lambda_0\right) / \lambda_0$ (so redshifts correspond to positive velocities). For $v_\text{shift}$ and $v_\text{width}$, the maximum velocity used for the integrations, $v_{max}$, is $\pm5000\,\text{km}\,\text{s}^{-1}$, while for $v_\text{skew}$, which is more sensitive to the line wings, this is $\pm10000\,\text{km}\,\text{s}^{-1}$. For Equation \ref{eq:widthEQ} the factor $2.35$ is used to ensure that, for a Gaussian profile, $v_\text{width}$ corresponds to the FWHM. Equations \ref{eq:centroidEQ} and \ref{eq:widthEQ} are similar to Equations 7 and 8 from \citet{jerkstrand2020properties} except that we here use the wavelength and spectral luminosity, rather than energy and count rates (as was suitable for gamma-ray lines). Skewness was not addressed in \citet{jerkstrand2020properties}, but is a new metric for nebular line profiles we introduce in this paper. 

The first moment, $v_\text{shift}$, diagnoses the bulk shift of the emission along the line of sight. This tells us how much more (or less) emission comes from the approaching hemisphere of the ejecta compared to the receding. Note that axisymmetric emissivisity distributions (e.g. jets) will still give zero $v_\text{shift}$ as long as the centre of the distribution is around zero line-of-sight velocity.

The second moment, $v_\text{width}$, measures the width of the line. Using the square of velocity shifts, it contains no information about line asymmetry per se, but probes the overall velocity scale of the emitting material. It does depend on the profile of the emission (versus velocity), so e.g. jets could be distinguished from other axisymmetric distributions.

The third moment, $v_\text{skew}$, using an odd power (as $v_\text{shift}$), again probes asymmetry. This time asymmetry with respect to the (possibly shifted) line centre, $v_\text{shift}$. E.g., a Gaussian (or other symmetric) line profile with a shifted line centre would have skewness zero. This could happen e.g. for a single (redshifted or blueshifted) blob dominating the emission. Thus, the skewness measures the degree of deviation from reflection symmetry of the line. The third power favors large $[v(\lambda) - v_\text{shift}]$ values, thus the sign of the skewness tends to indicate which line side is more extended (has a longer ``wing''). 


\subsection{Physical conditions}
In this section we will investigate two key physical properties of the model, namely the temperature  $T$ and electron fraction $x_e$. We will also study how these properties evolve over time. 

\subsubsection{Temperature evolution}
\begin{figure}
    \includegraphics[width=\columnwidth]{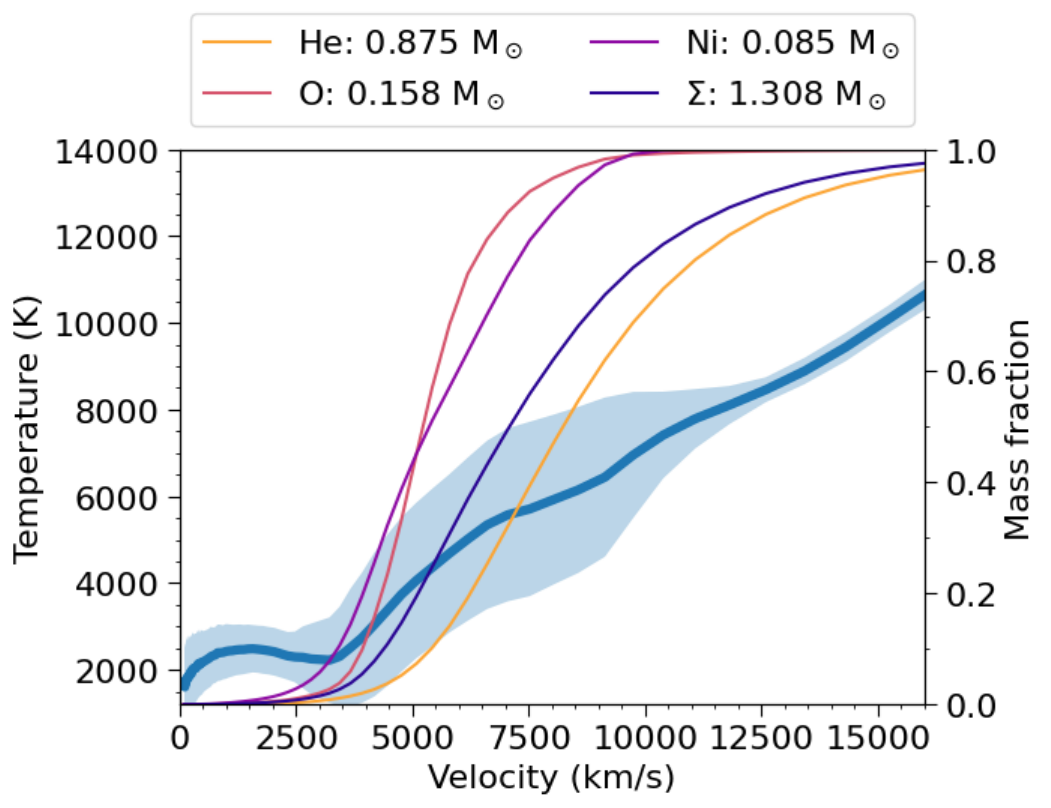}
        \caption{The angle-averaged temperature profile in the ejecta at 200d (thick blue line, left axis), with the shaded region corresponding to one standard deviation. The accumulative fractional masses of He, O, $^{56}$Ni and the total M$_{\text{ejecta}} (\Sigma)$ are also shown against the right axis. The figure showcases that the temperature variation with position angle, for a given radial velocity, is largest at intermediate velocities ($3,000-10,000$ km s$^{-1}$), where most of the metals are located.}
    \label{fig:HEC_33E_200day_T}
\end{figure}
In Figure \ref{fig:HEC_33E_200day_T} we showcase the angle-averaged temperature evolution together with the standard deviation at each velocity, for the HEC-33 model at 200 days post explosion. The figure showcases that the innermost regions ($v\leqslant3000\,\text{km}\,\text{s}^{-1}$) have a relatively low temperature of around $2000\,$K, with a standard deviation of $\sim500\,$K. 
Material at more intermediate velocities ($3000\leqslant v\leqslant9000\,\text{km}\,\text{s}^{-1}$) becomes, on average, progressively hotter the further out it is, but the spread in temperatures is also much larger for these radii $-$ reaching around $2000\,$K. At $v\geqslant9000\,\text{km}\,\text{s}^{-1}$ almost all of the ejecta is pure He, which results in much more spherical outer ejecta. This is reflected in the rapid drop of the temperature spread at these highest velocities, which in particular past $12000\,\text{km}\,\text{s}^{-1}$ are very small. The average temperature continues to rise  past the plotted range until $\sim20000\,\text{km}\,\text{s}^{-1}$, where the cells become extremely tenuous and the energy deposition by $\gamma$-rays becomes negligible and as a result the temperature drops (see also Section \ref{sec:Methods_code}). 

\subsubsection{Electron Fraction (x$_e$) evolution}
\begin{figure}
    \includegraphics[width=\columnwidth]{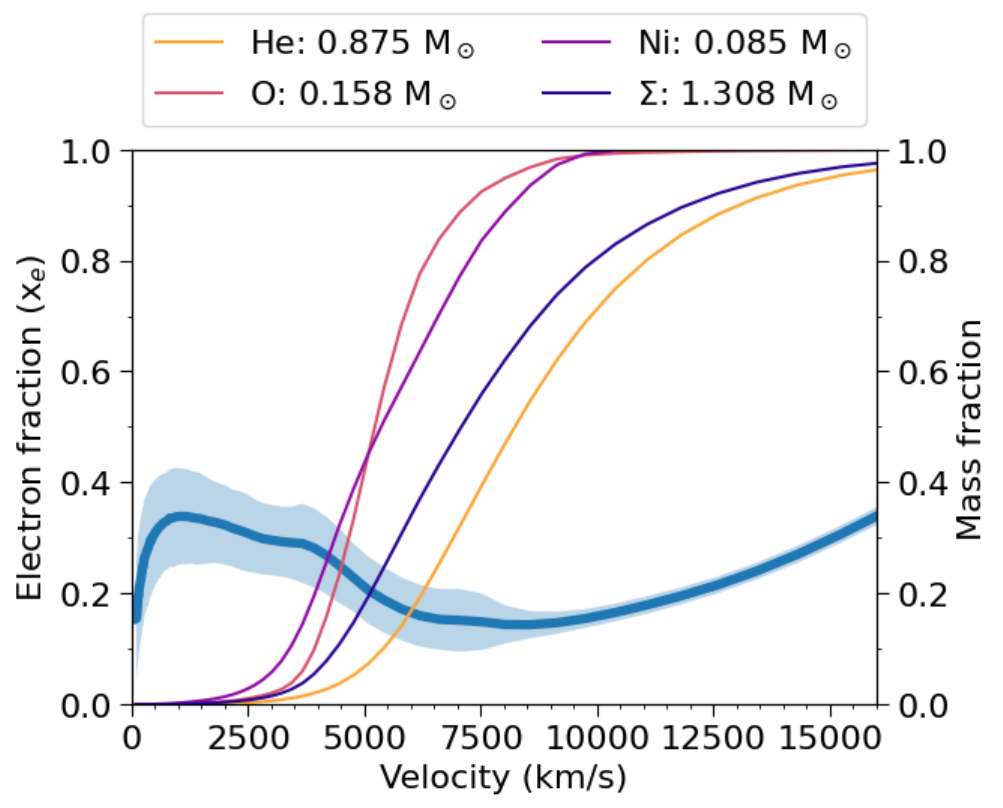}
    \caption{The angle-averaged electron fraction ($x_e$) profile in the ejecta at 200d (thick blue line, left axis), with the shaded region corresponding to one standard deviation. The accumulative fractional masses of He, O, $^{56}$Ni and the total M$_{\text{ejecta}} (\Sigma)$ are also shown against the right axis. The figure showcases that both low and intermediate velocity regions ($\lesssim 8,000$ km s$^{-1}$) have significant variation with position angle for the free electron fraction and thus the ionization structure.}
    \label{fig:HEC_33E_200day_xe}
\end{figure}
In Figure \ref{fig:HEC_33E_200day_xe} the angle-averaged electron fraction $x_e$ is shown together with the standard deviation across the HEC-33 model at 200 days post explosion. This fraction is calculated by counting all the free electrons in a cell and then dividing it by the total sum of all atoms and ions present in that cell. 

It can be observed that for $v\leqslant9000\,\text{km}\,\text{s}^{-1}$ the spread in $x_e$ is quite large for each velocity, with a sharp increase at the innermost radii followed by a gradual decent of the average electron fraction up to that point. For $v\geqslant9000\,\text{km}\,\text{s}^{-1}$ again the model becomes more spherical with almost pure He cells, and this results in the spread becoming much smaller. For these outer cells the electron fraction also starts to increase again, as the cells become less dense and thus recombination becomes less effective and more free electrons remain. This figure also shows that we are probably including enough ionization stages (neutral, singly ionized and doubly ionized) in our model, as even the most electron-rich cells have $x_e \ll 1$. This shows that the fraction of doubly-ionized elements is generally low and thus triple ionization will not come into play.

\subsubsection{Temporal Evolution}
\begin{figure}
    \includegraphics[width=\columnwidth]{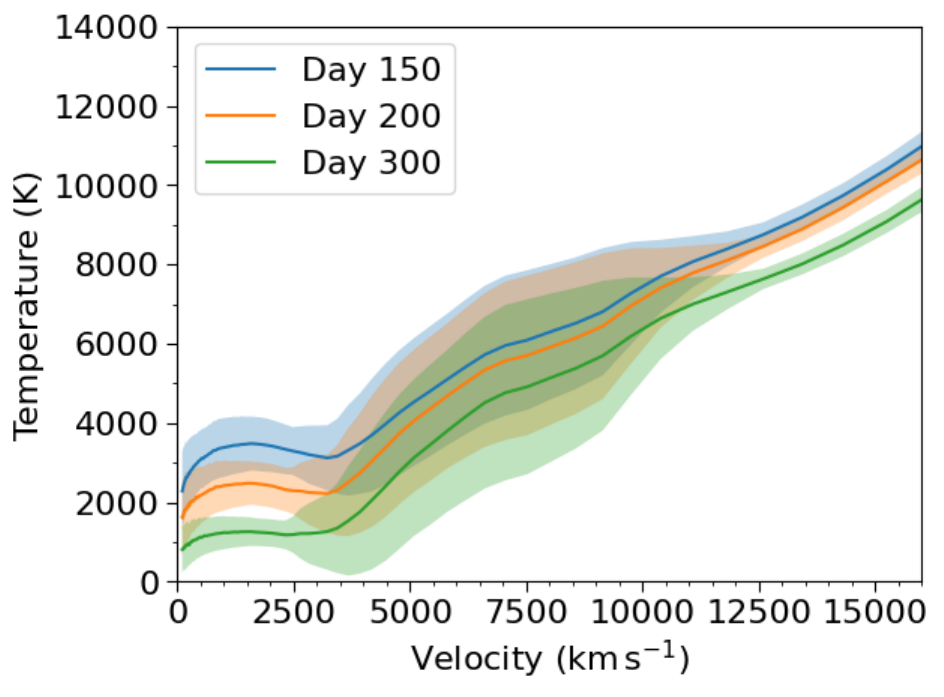}
    \includegraphics[width=\columnwidth]{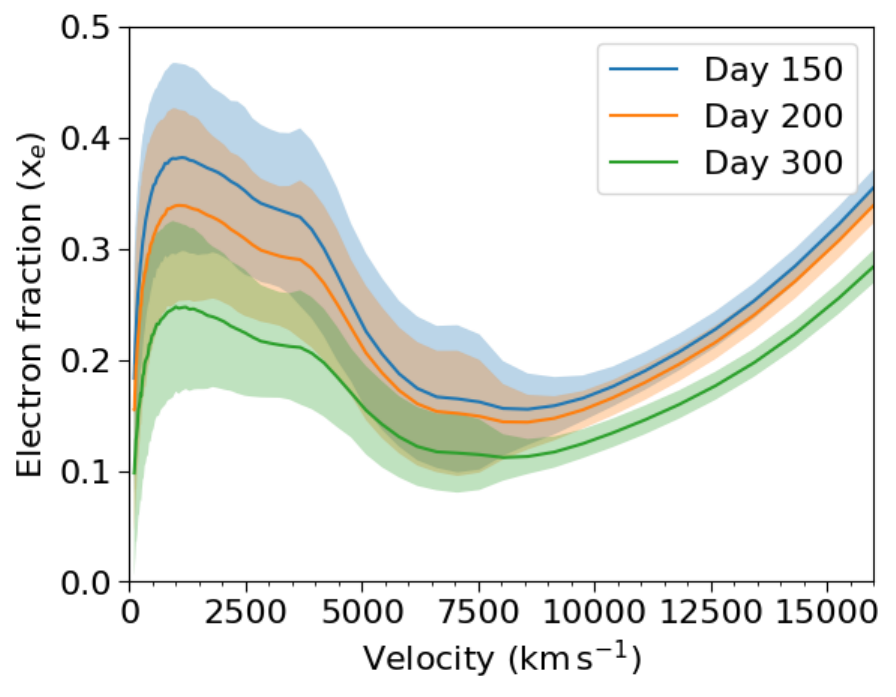}
    \caption{The temporal evolution of the  temperature profiles (top) and electron fraction profiles (bottom), for the HEC-33 model at  $150\,$days (blue), $200\,$days (orange) and $300\,$days (green). The total energy deposition by $\gamma$-rays is $2.24\times10^{40}$, $8.25\times10^{39}$ and $1.52\times10^{39}\,\text{erg}\,\text{s}^{-1}$, for each of the three epochs, respectively. The solid lines are the angle-averaged mean values, and the shaded regions correspond to one standard deviation.}
    \label{fig:HEC-33E_both_temporal}
\end{figure}
In Figure \ref{fig:HEC-33E_both_temporal} the temperature and electron fraction ($x_e$) distributions are shown for three different epochs ($150$, $200$ and $300\,$days). Overall, the general shapes of the distributions are similar at different epochs. One minor deviation is that for the innermost regions ($v\leqslant3000\,\text{km}\,\text{s}^{-1}$) the temperature towards later epochs seems to lose its 'bump' and just transitions to a flat slope. For the electron fractions such a flattening of the 'inner bump' does not take place between $150$ and $300\,$days. In each of the epochs, the fast, more spherical He-dominated ejecta remains at $v\geqslant9000\,\text{km}\,\text{s}^{-1}$ where the distributions become much narrower. 

The standard deviation decreases slightly at $v\leqslant2500\,\text{km}\,\text{s}^{-1}$ towards the later epochs for the temperature yet remains fairly constant at higher velocities, while for $x_e$ there is only a minor decrease with time. Across the time span shown here the total energy deposited from $\gamma$-rays drops by more than an order of magnitude between $150$ and $300\,$days (from $2.24\times 10^{40}\,\text{erg}\,\text{s}^{-1}$ to $1.52\times 10^{39}\,\text{erg}\,\text{s}^{-1}$), and although this clearly has an impact on both the temperature and $x_e$ curves it does not change their velocity profiles drastically. This indicates that we are set up in a good regime for our code where we do not suddenly encounter any problematic outliers.

\begin{figure*}
    \includegraphics[width=\textwidth]{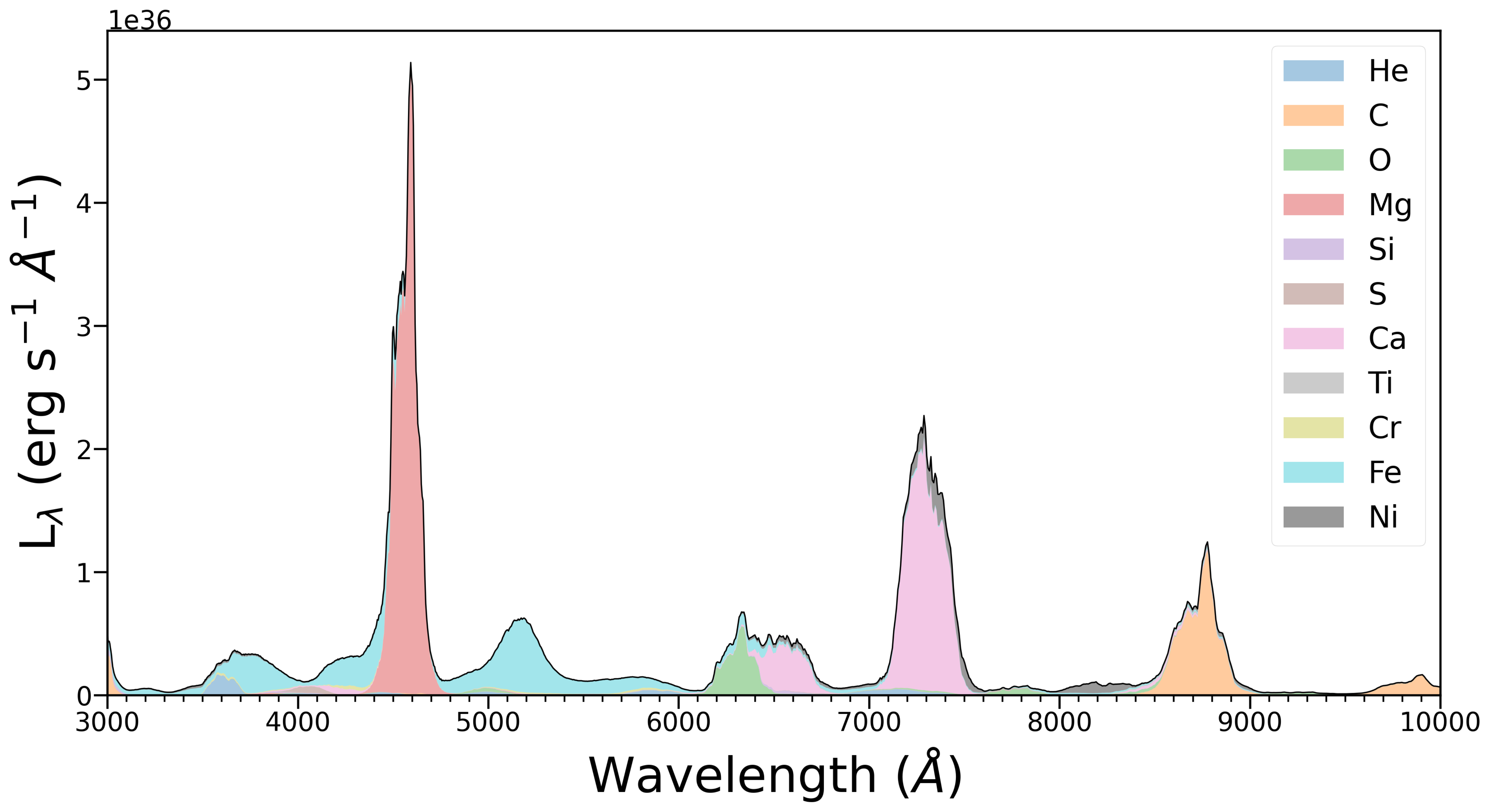}
    \caption{The optical model spectrum at $200\,$days for a particular viewing angle ($\theta=68^{\circ}$ and $\phi=27^{\circ}$ with respect to the axes as in Figure \ref{fig:exploModel3D}), with the contributions from the various elements coded by colour. The total spectrum is given by the black line. Only elements which create features in this wavelength range are included here. The viewing angle was not chosen for any particular reason.}
    \label{fig:HEC-33E_spectrum_per_el}
\end{figure*}
\subsection{Spectra at 200 days}
In Figure \ref{fig:HEC-33E_spectrum_per_el} we show the synthetic spectral output for the $3.3\,$M$_\odot$ Helium core model at 200 days in the wavelength range $3000-10000\,\angstrom$\footnote{This range is chosen somewhat arbitrarily, although we wanted to include a large range around the optical wavelengths which are usually where nebular phase spectra are taken. \texttt{ExTraSS} is not limited to these wavelengths, and can additionally be used for the near- and mid-infrared, where e.g. JWST can observe.}, for one particular viewing angle. The spectrum is color-coded for the different elements creating each of the emission features. The strongest features present are Mg I] $\lambda4571$, [Ca II] $\lambda\lambda7291,\,7323$, [C I] $\lambda8727$ and [O I] $\lambda\lambda6300,\,6364$. The bump of Fe-emission around $5200\,\angstrom$ is a blend of several Fe I and Fe II features, while the Ca-emission around $6500\,\angstrom$ has its origins in the [Ca I] $\lambda6573$ feature, which is a strong, low level feature. Notably, [O I] $\lambda\lambda6300,\,6364$, which is often classified as the strongest feature in Type-Ib observational spectra \citep[see e.g.][]{filippenko1997optical,taubenberger2009nebular,fang2022core,prentice2022oxygen}, is only the fourth strongest feature in our spectrum. One potential explanation for this is the lack of ionizing radiation field in our code $-$ Mg I and Ca I (and to a lesser extent, also C I) are easier to photoionize than O I, and are often heavily suppressed by photoionization \citep{jerkstrand2015late}.
This means that those states become overpredicted in this model lacking photoionization. Since Mg I and Ca I are very efficient coolers, overpredicting their populations will lead to a strong exaggeration of their emission, while damping the [O I] $\lambda\lambda6300,\,6364$ doublet which normally does most of the cooling. 

Some nebular phase spectra show a bump at $\sim6500\,\angstrom$, which can be a bit blended with the [O I] $\lambda\lambda6300,\,6364$ doublet, and the origins of this feature are contested between H$\alpha$ and [N II] $\lambda\lambda6548,\,6583$ \citep{patat1995late,jerkstrand2015late,fang2018origin}. We do not have either H nor N in our explosion model and as such do not obtain lines from these elements. We do, however, still obtain a bump around that wavelength coming from the earlier mentioned [Ca I] $\lambda6573$ feature. As explained above, it is likely that this feature's appearance is due to the lack of a radiation field in our model, as neutral calcium is fairly easy to photoionize. However, this might be indicating that there is yet another element (neutral Ca) which could contest with the red wing of the [O I] $\lambda\lambda6300,\,6364$ in some cases, in addition to the aforementioned H$\alpha$ and [N II] doublet.

\subsubsection{Line profiles and properties} 
\begin{figure*}
    \includegraphics[width=\textwidth]{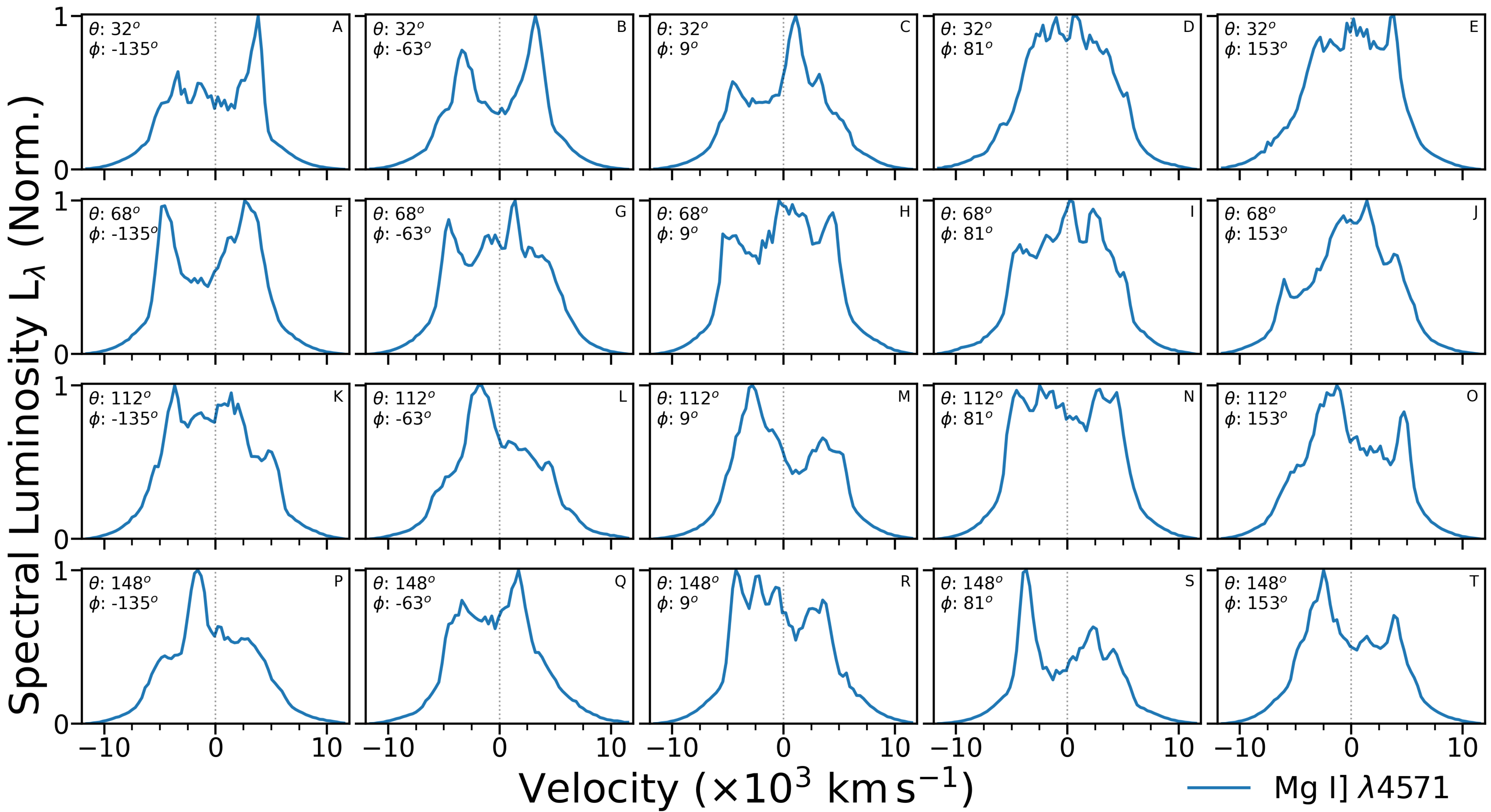}
    \caption{A display of 20 normalized line profiles of Mg I] $\lambda4571$ (taking just the contribution from magnesium to the plotted wavelength window). Each row displays line profiles for five different azimuthal angles $\phi$, $72^{\circ}$ apart, for a fixed polar angle (from top to bottom, $\theta=32^{\circ}$, $68^{\circ}$, $112^{\circ}$, and $148^{\circ}$). Each profile is normalized to its peak flux. The line profiles have been smoothed with a Gaussian with $R=1000$, a typical instrumental resolution.}
    \label{fig:HEC-33E_viewingbox}
\end{figure*}
With our $20\times20$ viewing angles we can investigate what sort of variations exist for line profile shapes depending on viewing angle. In Figure \ref{fig:HEC-33E_viewingbox} a display of the line profiles for 20 of these angles is shown, all focused exclusively on the Mg I] $\lambda4571$ feature. We choose this line because it is a singlet, is strong in our model, and often well observed. 

What can immediately be seen is the significant variation of line profiles with viewing angle $-$ there are double horned profiles (e.g. panels B, F), flat-top (N) and flat-top like (E, R), Gaussian-like (D), asymmetric single-peaked profiles (L, P), and several kinds of combinations between these. We can also see that viewing angle variation occurs both over polar and azimuthal directions, as no row nor column holds to one pattern. This is clearly a result of the fundamentally 3D, non-axisymmetric distribution of the (magnesium) ejecta. As these different profiles arise the same explosion model, this might indicate that different types of line profiles are not linked to specific kinds of SNe or progenitor stars, but may rather be a viewing angle effect. As we assume a globally optically thin ejecta, opposite viewing angles will give mirrored line profiles $-$ it should be noted that Figure \ref{fig:HEC-33E_viewingbox} holds no exact mirror pairs.

\begin{figure*}
    \includegraphics[width=.495\textwidth]{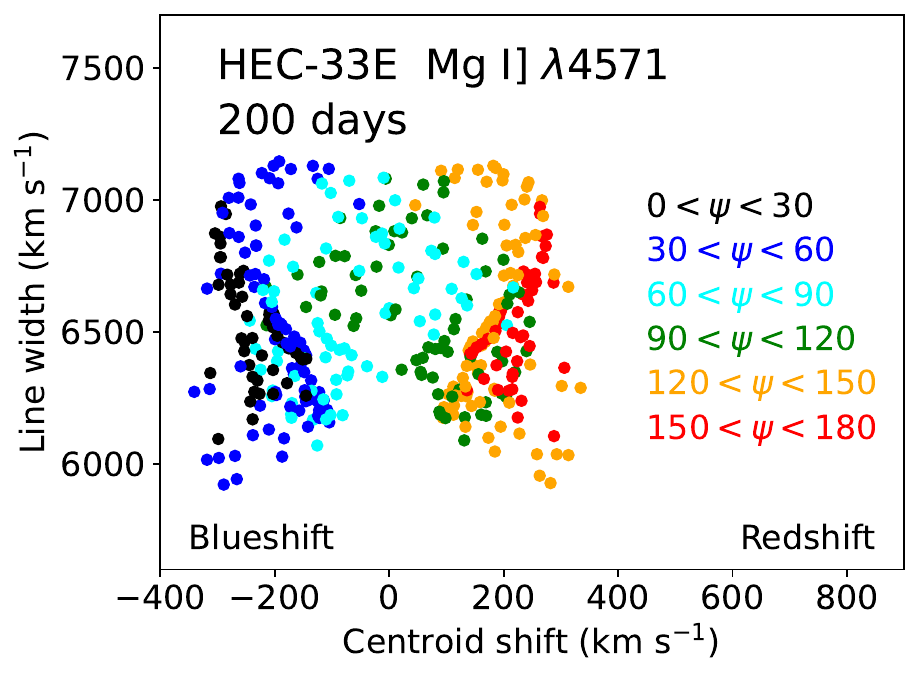}
    \includegraphics[width=.495\textwidth]{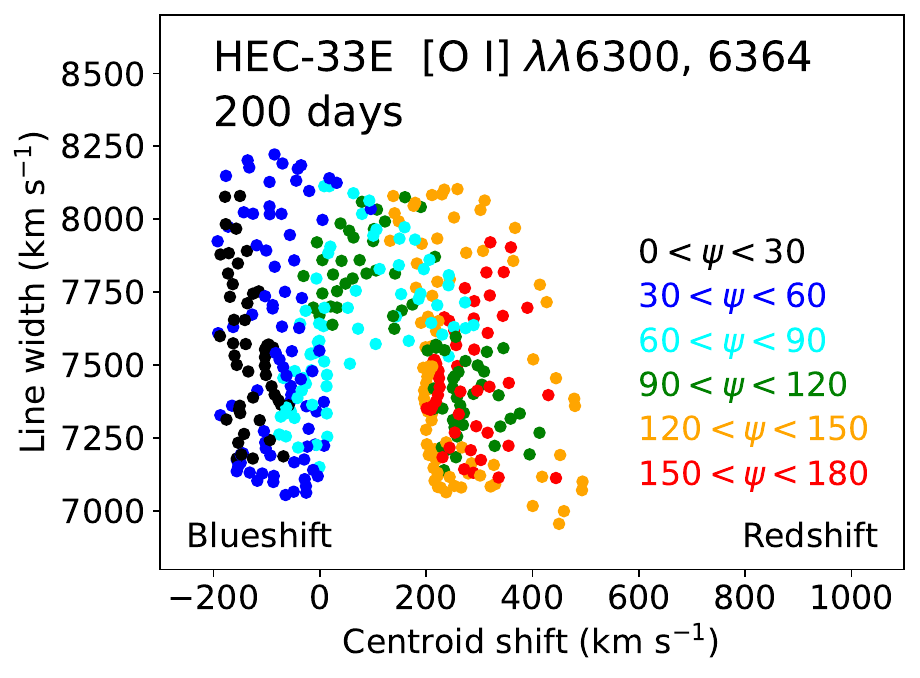}  
    \includegraphics[width=.495\textwidth]{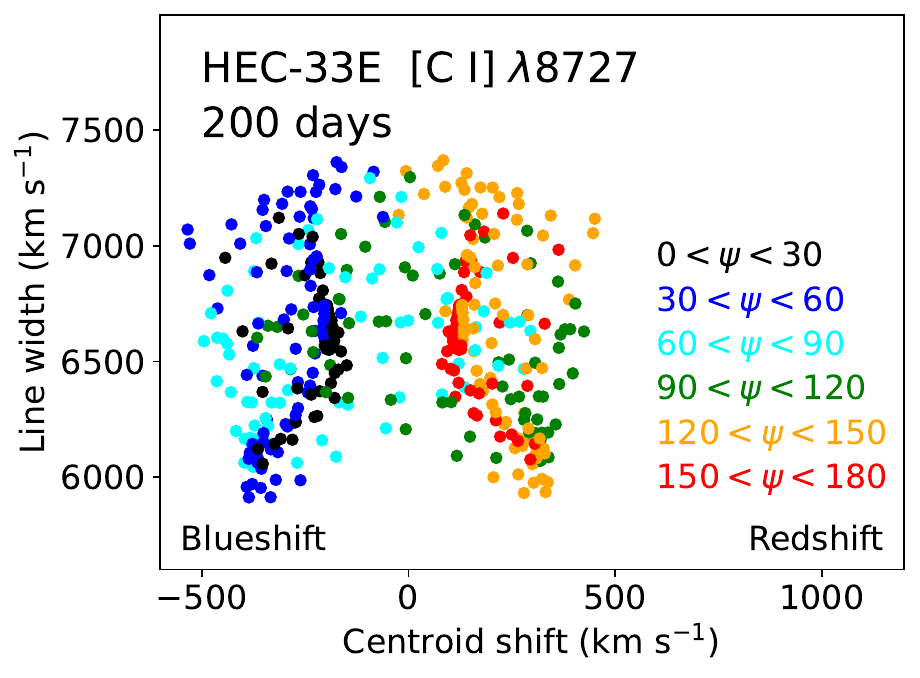}
    \includegraphics[width=.495\textwidth]{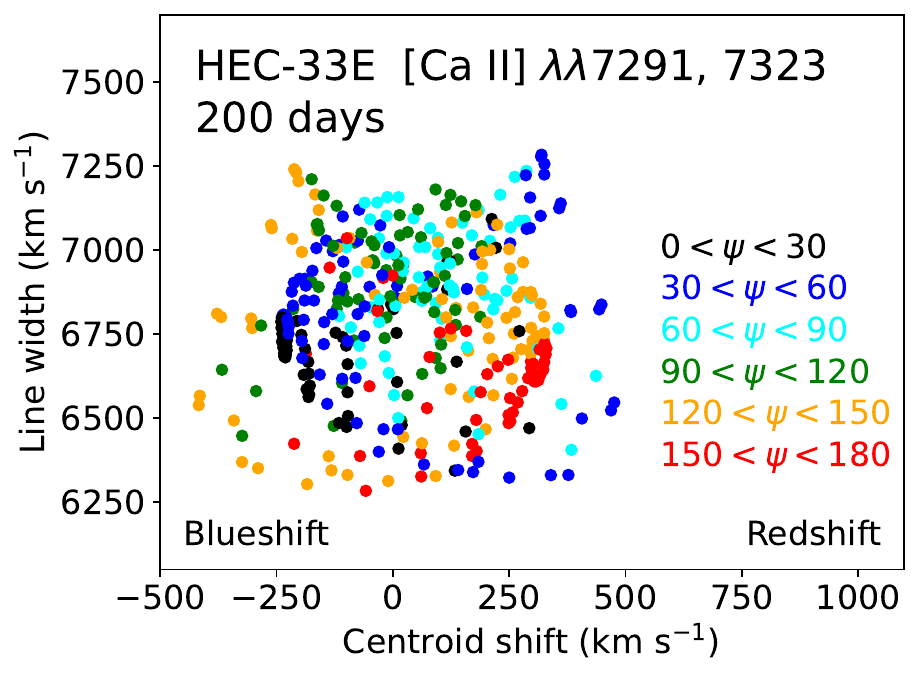} 
    \caption{The centroid line shifts (x-axis) and (FWHM) line widths (y-axis) of the singlet features Mg I] $\lambda4571$ (top left) and [C I] $\lambda8727$ (bottom left), and the doublet features [O I] $\lambda\lambda6300,\,6364$ (top right) and [Ca II] $\lambda\lambda7291,\,7323$ (bottom right), for the 200 day HEC-33 model. The centroid shifts (Eq. \ref{eq:centroidEQ}) are relative to the rest wavelength of the singlet features and relative to the transition-strength weighted rest wavelength for the doublet features, being $6316\,\angstrom$ for [O I] and $7304\,\angstrom$ for [Ca II]. The line widths are calculated with Equation \ref{eq:widthEQ}. The $\Psi$ angle is the angle between the direction vector to the viewer and the neutron star motion vector, i.e. the black points (small $\Psi$ values) correspond to viewing angles where the NS is moving almost directly towards the observer, while the red points (large $\Psi$) correspond to viewing angles for which the NS moves almost radially away from the observer. A positive centroid shift value corresponds to a redshifted centroid.}
    \label{fig:lineprofiles_all}
\end{figure*} 

In Figure \ref{fig:lineprofiles_all} we show line shifts and widths for the four strongest features in our synthetic spectrum: the singlets Mg I] $\lambda4571$ and [C I] $\lambda8727$ (left column) and the doublets [O I] $\lambda\lambda6300,\,6364$ and [Ca II] $\lambda\lambda7291,\,7323$ (right column). The line profiles are coloured by the angle $\Psi$ between viewing direction and neutron star kick vector, with the black dots corresponding to well-aligned viewing angles and the red dots to the anti-aligned viewing angles. 

The singlet features have FWHM values of $\sim6500\pm700\,\text{km}\,\text{s}^{-1}$ and centroid shift values between $-400$ to $+400\,\text{km}\,\text{s}^{-1}$ (Mg I]) and $-500$ to $+500\,\text{km}\,\text{s}^{-1}$ ([C I]). For these singlet line profiles, centroids which are redshifted correspond mostly to viewing angles with $\Psi > 90^\circ$ while blueshifted centroids correspond to $\Psi < 90^\circ$, indicating that these elements have similar bulk motions. Interestingly, however, the centroid shifts for viewing angles closest to (anti-)alignment with the neutron star kick (black and red points) for the Mg I] feature are exclusively among the viewing angles with the highest centroid shift, while this is not the case for the [C I] line. Instead, for the [C I] feature they are among the more narrow profiles while also displaying a spread in centroid shift values. So while the carbon and magnesium emissivity distributions are partially overlapping, there are still differences. 

For singlet features, centroid shifts are expected to be symmetric for anti-aligned observers: if the centroid is offset by any particular value $X$ along viewing angle $Y$ then, for an optically thin nebula, the opposite viewing angle $Y_\text{opp}$ will observe an offset of $-X$. Similarly, both $Y$ and $Y_\text{opp}$ should observe the same line width $Z$, and thus we get a distribution mirrored on the zero shift line. This is also seen, although there is a small offset from zero in Figure \ref{fig:lineprofiles_all} which appears because the rest-wavelengths of the features are not exactly in the middle of a wavelength bin. 

The other two strongest features in Figure \ref{fig:HEC-33E_spectrum_per_el} are the doublet features of [O I] $\lambda\lambda6300,\,6364$, which has a $3:1$ contribution ratio to the overall feature, and [Ca II] $\lambda\lambda7291,\,7323$ which has a $3:2$ contribution ratio: these two lines originate from different excitation levels which have the same transition strength but different statistical weights ($g=6$ and $g=4$, respectively). Both of these relations are under the assumption that the nebula is optically thin, and for the computed [Ca II] level ratios we find good agreement to this expected $3:2$ ratio. For the oxygen lines, the optically thin limit should occur somewhere between $100-200\,$days for stripped envelope SNe \citep{jerkstrand2015late} so this assumption should also be valid for our $200\,$day spectrum, resulting in a 3:1 ratio between the two lines in our model. 

The line shifts and widths of these two doublets is shown in Figure \ref{fig:lineprofiles_all} (right column). It can be seen that full symmetry is not achieved for the doublets. Apart from asymmetry, the [O I] doublet also has a $\sim$ 100 km s$^{-1}$ offset from the zero shift, corresponding to 2 \AA. This may be explained by a weighted rest wavelength deviating somewhat from the optically thin value ($6316\,\angstrom$), and/or some flux falling outside our integration interval.


The FWHM of the [Ca II] feature is roughly mirrored, but there are certain differences between the redshifted and blueshifted sides. The viewing angles of [Ca II] do not match the pattern displayed by the other three lines, as for [Ca II] there is no clear separation between the aligned (black) and anti-aligned (red) viewing angles with respect to the neutron star. For any set of $\Psi$, both a redshifted and a blueshifted centroid shift can be found, as well as variations of the FWHM of $\sim800\,\text{km}\,\text{s}^{-1}$. Only the most strongly (anti-)aligned viewing angles are relatively grouped together, which might be indicating that there is a kind of a sheet of [Ca II] emitting material which is moving in nearly the same direction as the neutron star. However it is not a very strong effect and hence it is only picked up by these angles where $\Psi\leqslant 30^\circ$ or $\Psi\geqslant150^\circ$, while for other angles the velocity shifts caused by this sheet get washed out amongst the other [Ca II] emitting regions, leading to this non-split but mirrored feature. This deviation from the pattern displayed by the other profiles might be because the [Ca II] feature is created from transitions from the third and second excited state to the ground state, respectively. If the balance between these two levels varies across the ejecta then different viewing angles will pick up different patterns than for the singlet lines (or [O I] which is two transitions from the same upper level) which could lead to the changes described here. 

Putting the 20 angles from Figure \ref{fig:HEC-33E_viewingbox} into the perspective of Figure \ref{fig:lineprofiles_all} (top left panel), ($\Psi=0^\circ$ occurs at $\approx\theta=152^\circ$, $\phi=-4^\circ$), there is a mashing of the different viewing angles with respect to the NS. This variation leads to deviations between each column and row on what the centroid shift is for each angle. On the other hand, for the FWHM values it can be seen that the angles shown here indeed display FWHM$\sim6500\pm700\,\text{km}\,\text{s}^{-1}$ although some still display some quite narrow features on one or occasionally both sides. 

The middle two rows in Figure \ref{fig:HEC-33E_viewingbox} mostly correspond to the green ($90^\circ<\Psi<120^\circ$) and light blue ($60^\circ<\Psi<90^\circ$) dots in Figure \ref{fig:lineprofiles_all} and we can indeed see that the centroids for those line profiles do not strongly favour either a redshift or blueshift. Just as with the top and bottom rows though, also for the FWHM of these profiles display large variations with $\phi$ which matches the spread seen in Figure \ref{fig:lineprofiles_all}.

\subsubsection{Skewness}
\begin{figure*} 
    \includegraphics[width=.495\textwidth]{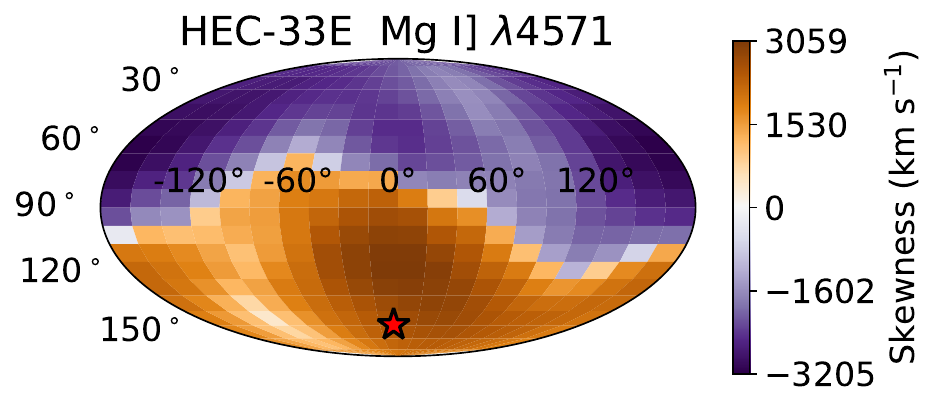}
    \includegraphics[width=.495\textwidth]{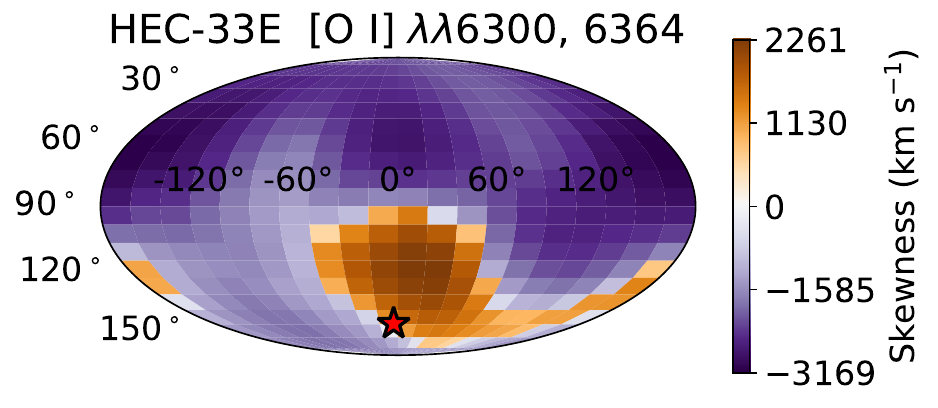}
    \includegraphics[width=.495\textwidth]{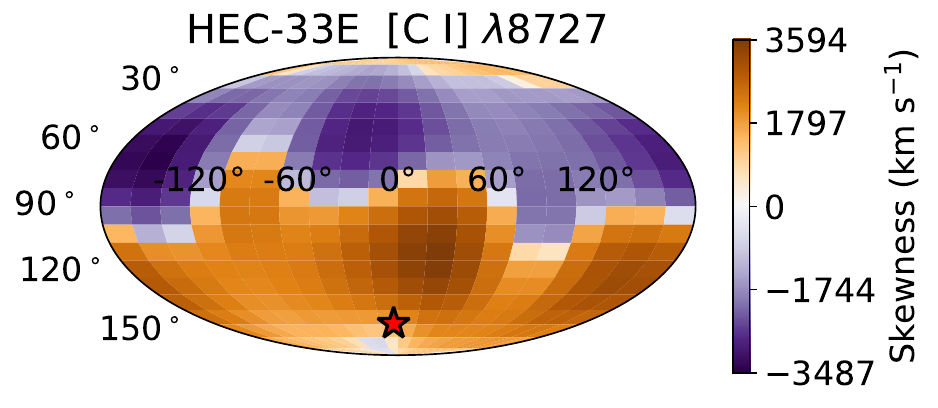}
    \includegraphics[width=.495\textwidth]{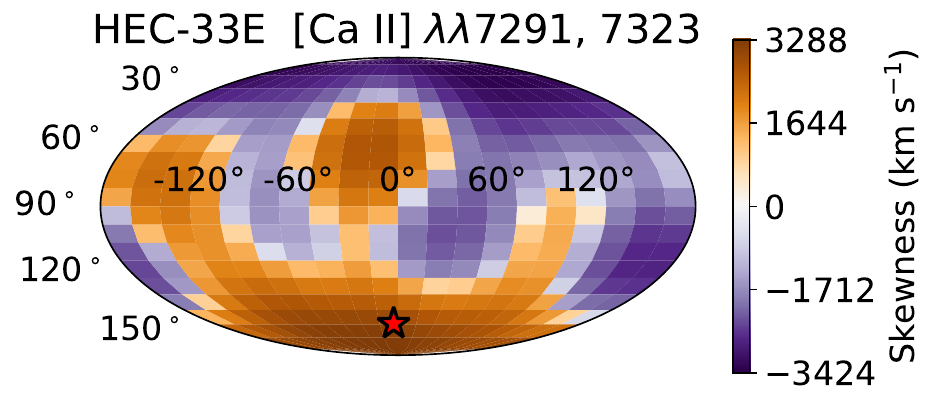}
    \caption{Mollweide displays for the skewness (Eq. \ref{eq:skewEQ}), at 200d, of the four main features Mg I] $\lambda4571$ (top, left), [O I] $\lambda\lambda6300,\,6364$ (top, right), [C I] $\lambda8727$ (bottom, left), [Ca II] $\lambda\lambda7291,\,7323$ (bottom, right). A positive skewness indicates a stronger redshifted tail compared to the blue one, and vice-versa. It should be noted that each panel has its own colour scale, although in all cases orange/brown colours are the most positively skewed profiles, white colours the most neutral and purple the most negatively skewed. The neutron star motion direction is marked with the red star.}
    \label{fig:skewness_combi}
\end{figure*}
Beyond looking at the primary line measurements (shifts and widths), it is also interesting to investigate variations of the skewness of the line profiles for different viewing angles and elements. To calculate the skewness we use Equation \ref{eq:skewEQ}. The skewness can be an interesting metric because it can indicate if there are strong tails of material in particular directions which can hint at strong asymmetries in the ejecta, in particular for singlet lines as they should be mirrored under opposite viewing angles for an optically thin nebula. 

In Figure \ref{fig:skewness_combi} we show the skewness of the four strongest lines in our model in a Mollweide projection for all viewing angles. As with centroid, in the absence of non-local radiative transfer, skewness is expected to have reflection symmetric for opposite side observers, with the same magnitude but opposite signs. We can see that the two singlet lines, Mg I] $\lambda4571$ and [C I] $\lambda8727$, indeed follow this behaviour quite closely $-$ most of their viewing angle pairs  (i.e. $180^\circ$ apart in the azimuthal direction and mirrored on the polar angle) have the same absolute values and mirrored in sign. Their pattern of positive and negative skewness is also quite similar, again, just as in Figure \ref{fig:lineprofiles_all} for the line widths and centroid shifts. It is mostly the viewing angles close to the NS angle which obtain a positive skewness, which implies that these are profiles with strong wings on their red emission side, although for [C I] there is an intriguing flip of signs at the extreme polar angles indicating a strong influence of emitting [C I] material at quite a close angle to the NS. 

For the doublet features, the picture is a bit more complicated. Doublet features are inherently asymmetric also for symmetric components, due to the different strengths of the components. For symmetric components with a FWHM of $6000\,\text{km}\,\text{s}^{-1}$, the  [Ca II] doublet has an intrinsic skewness of $v_\text{skew}\approx +400\,\text{km}\,\text{s}^{-1}$, while the [O I] doublet has an intrinsic skewness $v_\text{skew}\approx +800\,\text{km}\,\text{s}^{-1}$. 

The [Ca II] $\lambda\lambda7291,\,7323$ feature here also trends towards the mirror-effect for opposite viewers as in particular the $\theta\leqslant30^\circ$ viewing angles result in the most negatively skewed profiles, while the viewing angles with $\theta\geqslant150^\circ$ give the most positively skewed profiles. The two positively skewed bubbles at $\theta\approx75^\circ$, $\phi\approx-150^\circ$ and $\theta\approx60^\circ$, $\phi\approx-15^\circ$ also have negatively skewed mirrors at $\theta\approx110^\circ$, $\phi\approx-30^\circ$ and $\theta\approx130^\circ$, $\phi\approx150^\circ$, respectively) which indicates that for this doublet the mirror-assumption holds. However, the overall skewness pattern is a completely different overall pattern than for the two singlet lines. Instead, the regions which are most positively skewed for the singlet lines find negatively skewed [Ca II] profiles (e.g. at $\theta\approx120^\circ$, $\phi\approx30^\circ$), which is a strong indicator that the [Ca II] emission does not trace the Mg I] or [C I] emission. There are, however, viewing angles where the three features all have (strongly) negatively skewed profiles so for certain viewing angles these three profiles might look quite similar, such as most viewing angles where $\theta\leqslant30^\circ$. 

The differences between the skewness patterns for [Ca II] doublet compared to Mg I] and [C I] singlets might, to some extent, have its origin in the doublet nature of the [Ca II] feature. With the non-equal emission in the two components, at 3:2, the [Ca II] doublet has an intrinsic asymmetry in its combined profile (see above). However, the most extreme values for $v_\text{skew}$ in the [Ca II] doublet are very comparable to the singlet ranges, and an order of magnitude larger than the intrinsic value of this doublet, indicating that the [Ca II]-skewness per viewing angle is dominated by the asymmetries in the ejecta here. 

The [O I] doublet has an intrinsic skewness of $v_\text{skew}\approx +800\,\text{km}\,\text{s}^{-1}$ (see above), but in Figure \ref{fig:skewness_combi} we mostly see negative values instead. This is then not caused by the doublet nature, but may be due to to contributions from weaker lines (e.g. the 5s5S$\star$ to 3p5P at $\lambda\lambda 6453,\,6454,\,6456$ transition) which contaminate the wings of the [O I] doublet.

For three of the features in Figure \ref{fig:skewness_combi}, the maximum values for the negatively and positively skewed profiles are very similar $-$ but not for [O I], where the most negatively skewed value is roughly $50\%$ larger than the most positive skewness. This might be caused by the relatively high intrinsic skewness of the [O I] doublet, although we do have a stronger negative component rather than positive.

\subsection{Singlets $\&$ Doublets}
\begin{figure}
    \includegraphics[width=\columnwidth]{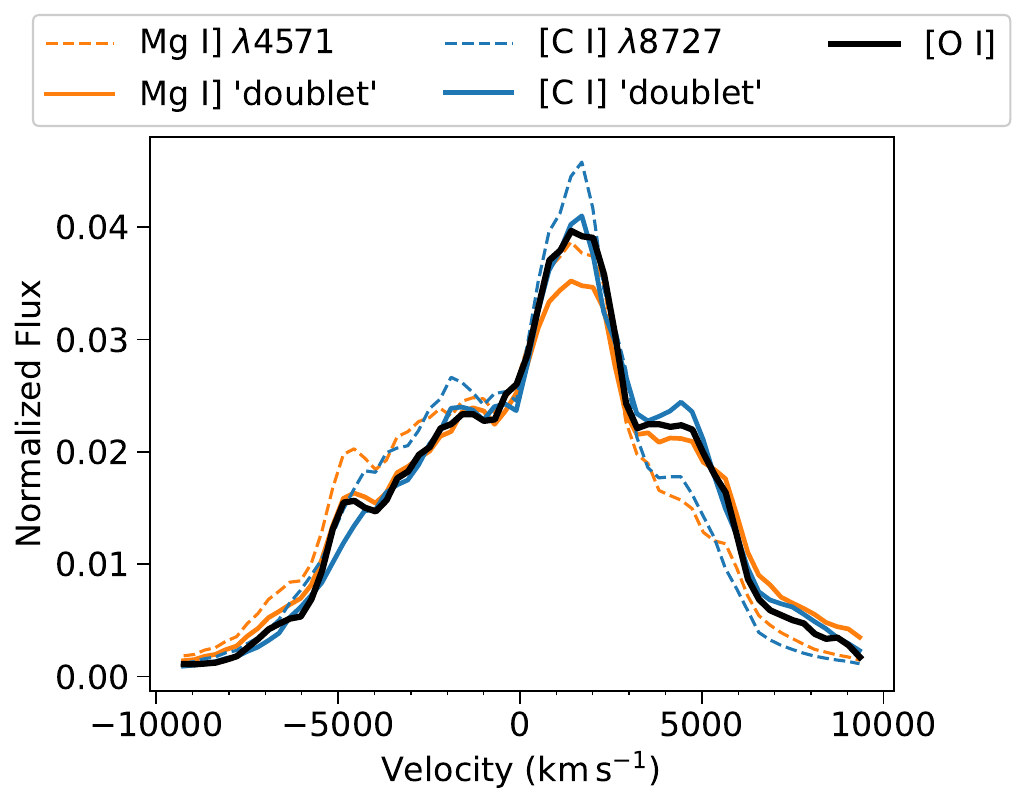}
    \caption{A comparison of the [O I] $\lambda\lambda6300,\,6364$ doublet line profile (solid black) to the Mg I] $\lambda4571$ (dashed orange) and [C I] $\lambda$8727 (dashed blue) singlet lines. The plot is for the same viewing angle as in Fig. \ref{fig:HEC-33E_spectrum_per_el}. As additional comparison, mock doublet features of the Mg I] and [C I] profiles are shown (bold lines) where a second feature is added with the same velocity offset (3047 km/s) and emission ratio (1/3) as the 6364 to 6300 Å emission, to mimic the behaviour of the [O I] $\lambda\lambda6300,\,6364$ doublet. For all doublets, $v=0$ here corresponds to the rest wavelength of the blue feature.
    }
    \label{fig:OMGC_linefeature}
\end{figure}
One of the notable features of Figure \ref{fig:HEC-33E_spectrum_per_el} is that the line profiles of Mg I] $\lambda4571$, [C I] $\lambda8727$ and [O I] $\lambda\lambda6300,\,6364$ look pretty similar in shape at first glance. From Figure \ref{fig:lineprofiles_all}, left panels, it can additionally be seen that the line widths and centroid shifts for Mg I] and [C I] display somewhat similar distributions and have a similar range of values. The blueshifted side ($\Psi < 90^\circ$) of the [ O I] feature in Figure \ref{fig:lineprofiles_all} also matches these to some extend. In Figure \ref{fig:OMGC_linefeature} we show each of these features centered at their respective rest wavelengths (i.e. $v=0\,\text{km}\,\text{s}^{-1}$ corresponds to $4571\angstrom$ for Mg I], to $8727\angstrom$ for [C I] and $6300\angstrom$ for [O I] here), at a particular viewing angle. It can be noted that the peaks of the [O I] doublet (black) and Mg singlet (dashed orange) align very well, but for the wings ($|v|\gtrsim 5000\,\text{km}\,\text{s}^{-1}$) the Mg feature has a stronger blue component and a weaker red wing, comparatively. The [C I] singlet (dashed blue) has a decent agreement for the centre of the line but also here in the wings the blue component is relatively strong and the red component relatively weak. Each of the curves is normalized to the sum of its own spectral luminosity. 

However, here we are comparing two singlet features to a doublet feature, so part of any differences is simply due to this. To make a more informative comparison, we also created mock Mg I] and [C I] doublets to see whether these line profiles match better to the [O I] doublet. These mock doublets were created by taking the singlet line and adding $33\%$ of that feature but redshifted by (64/6300 * $\lambda_\text{rest}$), as that is the difference in line strength and location for the [O I] doublet under the optically thin assumption. These mock doublets are also shown in Figure \ref{fig:OMGC_linefeature} with the bold lines (orange for Mg I], blue for [C I]). 

It can be seen that for the magnesium mock-doublet, the peak line flux region in fact matches less well to the [O I] doublet, but outside this narrow peak region (with $1500\leqslant v\leqslant2500\,\text{km}\,\text{s}^{-1}$) it matches very well, better than the original singlet Mg I] profile. For the carbon mock-doublet it can be seen that also in the line peak region the agreement with the [O I] doublet is very good. This is a strong indicator that, at least for this viewing angle, the emissivity distributions of these elements are co-spatial to a significant degree, in particular carbon and oxygen. 

Figure \ref{fig:OMGC_linefeature} only displays the comparison for one viewing angle (the same viewing angle as in Figure \ref{fig:HEC-33E_spectrum_per_el}), and for the other angles the level of agreement might be different. In order to test this we performed a $\chi^2$-test between the Mg I], doublet-Mg I], [C I] and doublet-[C I] line profiles against the [O I] doublet. We performed this test against the normalized line profiles.

\begin{figure}
    \includegraphics[width=\columnwidth]{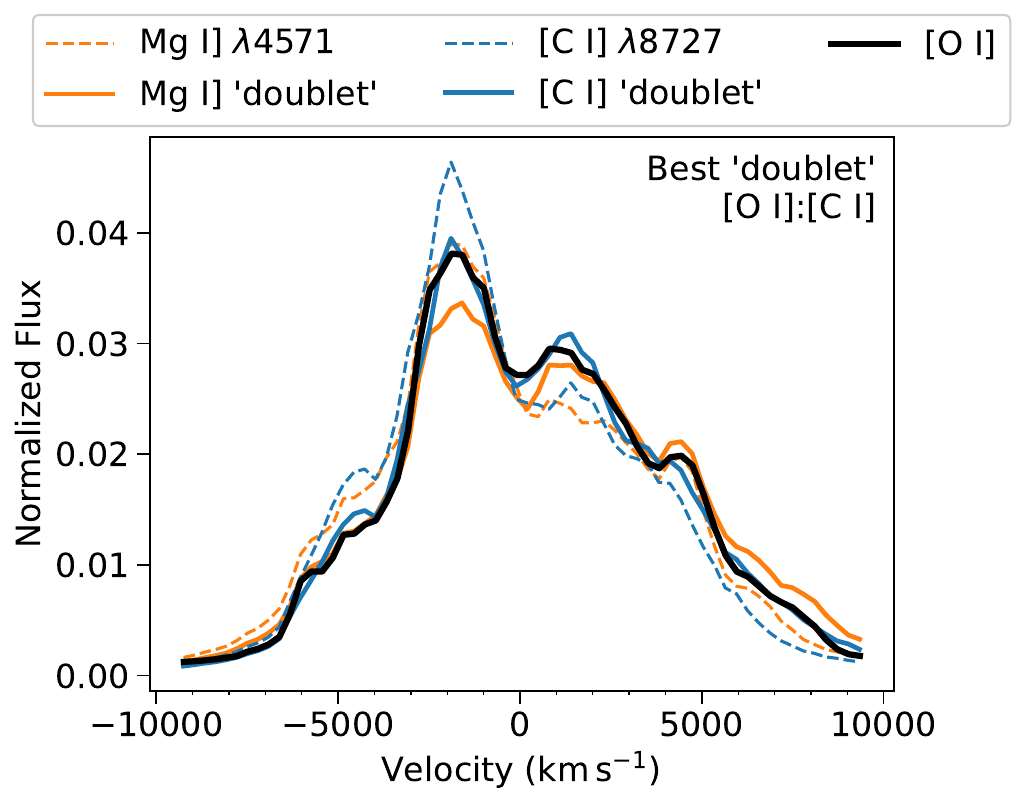}
    \includegraphics[width=\columnwidth]{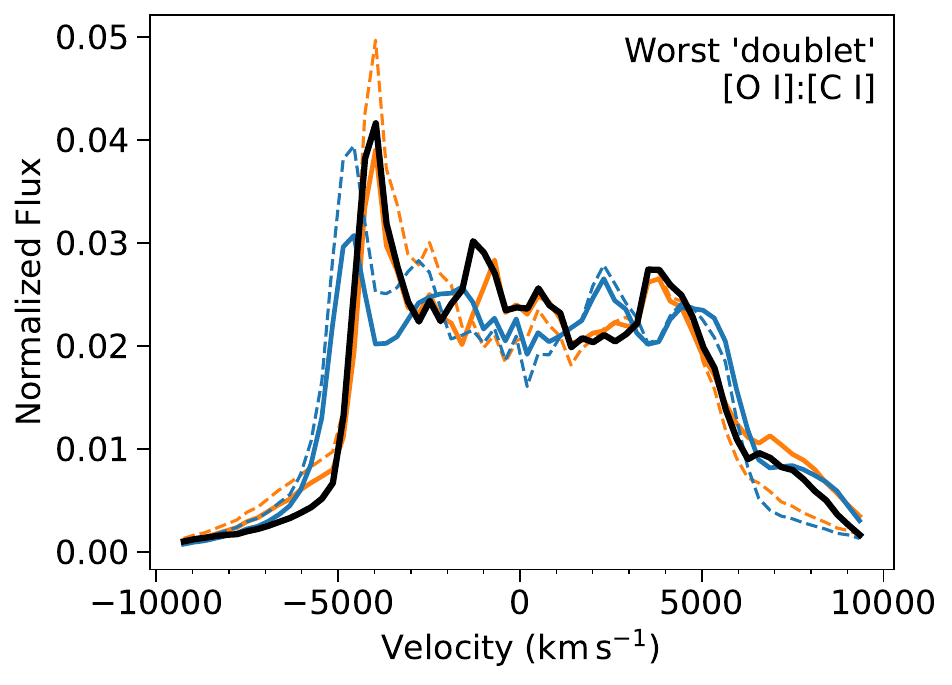}
    \caption{The [O I] $\lambda\lambda6300,\,6364$ doublet (solid black) profile compared to Mg I] $\lambda4571$ (orange dashed) and the mock Mg I] doublet (orange bold), and the [C I] $\lambda8727$ (blue dashed) and mock [C I] doublet (blue bold) profiles, for the particular angles $\theta=112^\circ$, $\phi=-63^\circ$ (top panel) and $\theta=139^\circ$, $\phi=45^\circ$ (bottom panel): these are the two viewing angles with the best and worst $\chi^2$ values, respectively, for the [C I] mock doublet compared to the [O I] doublet.}
    \label{fig:Cdoublet_match_VAbestworst}
\end{figure} 

In Figure \ref{fig:Cdoublet_match_VAbestworst} we showcase the line profile comparison for two different viewing angles, namely the angles for which the $\chi^2$ value for the [C I]-doublet vs [O I] doublet test comes out the best (top) and the worst (bottom). For the viewing angle in the top panel, the blue side of both singlet profiles (dashed lines) agrees quite well already with the [O I] doublet. For the red side it is clear that the singlet profiles do not capture the [O I] profile well, and the Mg I] mock doublet (bold orange) compensates too much and overpredicts the line profile on the red side for $v>5000\,\text{km}\,\text{s}^{-1}$. The [C I] mock doublet (bold blue) however, succeeds really well in matching the entirety of the original [O I] profile. With a value of $\chi^2=4.8\times10^{-3}$ (normalized as deviation from the [O I] doublet) between the mock-[C I] doublet and the [O I] doublet, this mock-[C I]-doublet : [O I]-doublet has the lowest $\chi^2$ for any of the four line profiles compared to the [O I] doublet across all viewing angles. 

In the bottom panel of Figure \ref{fig:Cdoublet_match_VAbestworst} we instead showcase the viewing angle where the line profiles of [O I] : [C I]-doublet have the worst agreement, according to the $\chi^2$ value (0.12, normalized as deviation to the [O I] profile). For the C-lines in particular, there is barely any improvement between the singlet line and doublet line fits to the [O I] profile while for the Mg-profiles the mock-doublet does actually match relatively well ($\chi^2$ for Mg I] doublet : [O I] is at 0.02 for this viewing angle). For the [C I]-doublet however, it overpredict the blue wing significantly, while it additionally struggles to capture the center part of the [O I] profile properly creating a small bump exactly where the [O I] does not display one. Furthermore, the [C I] line profile is somewhat broader than the other two elements for this viewing angle $-$ which is particularly interesting because the angle between these two viewing angles is quite close to $90^\circ$, but only one of the two mock doublets ([C I]) drastically changes in how well it fits to the [O I] doublet in line profile. 

\subsection{Mass slices} \label{ssec:slices}
\begin{figure*}  
    \includegraphics[width=.33\textwidth]{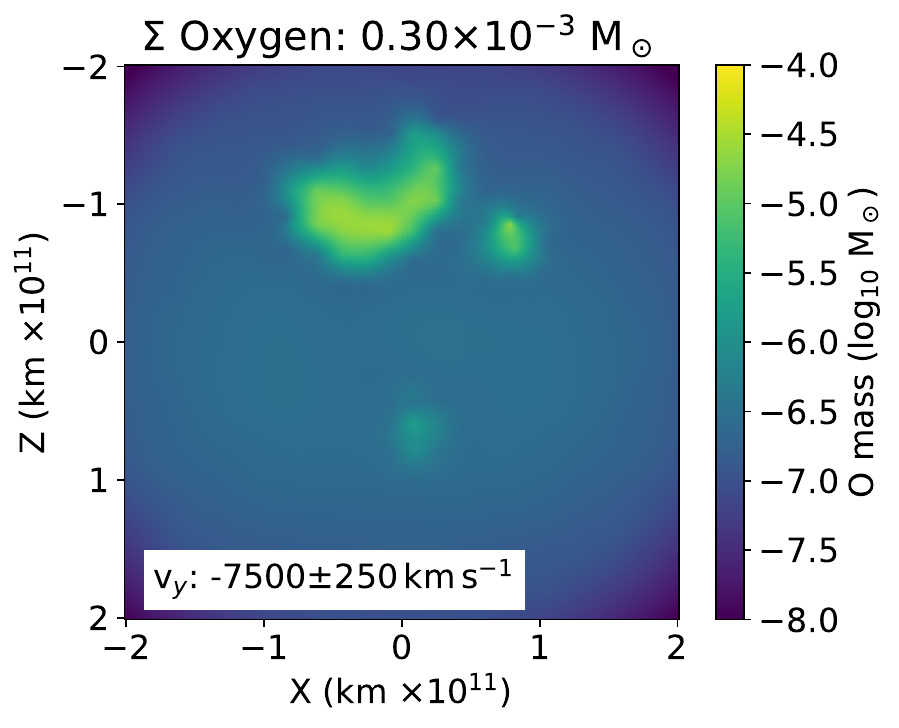}
    \includegraphics[width=.33\textwidth]{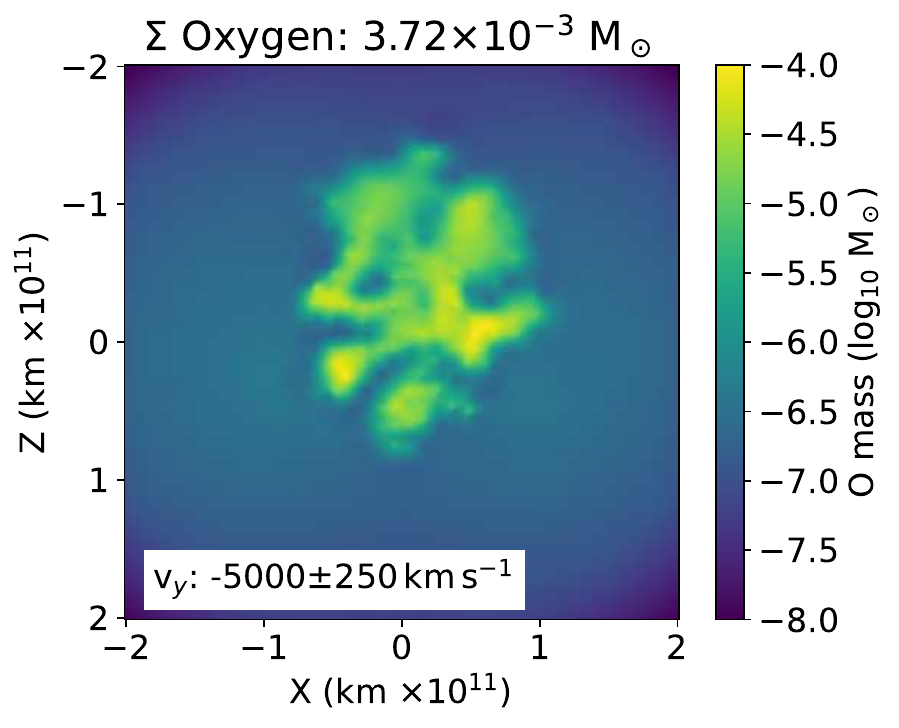}
    \includegraphics[width=.33\textwidth]{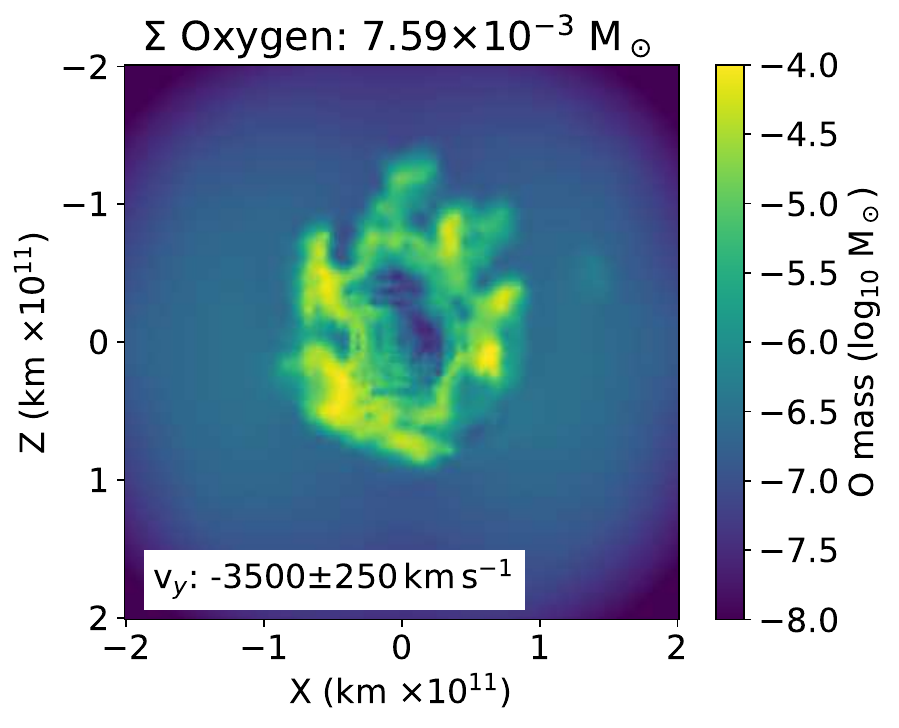}
    \includegraphics[width=.33\textwidth]{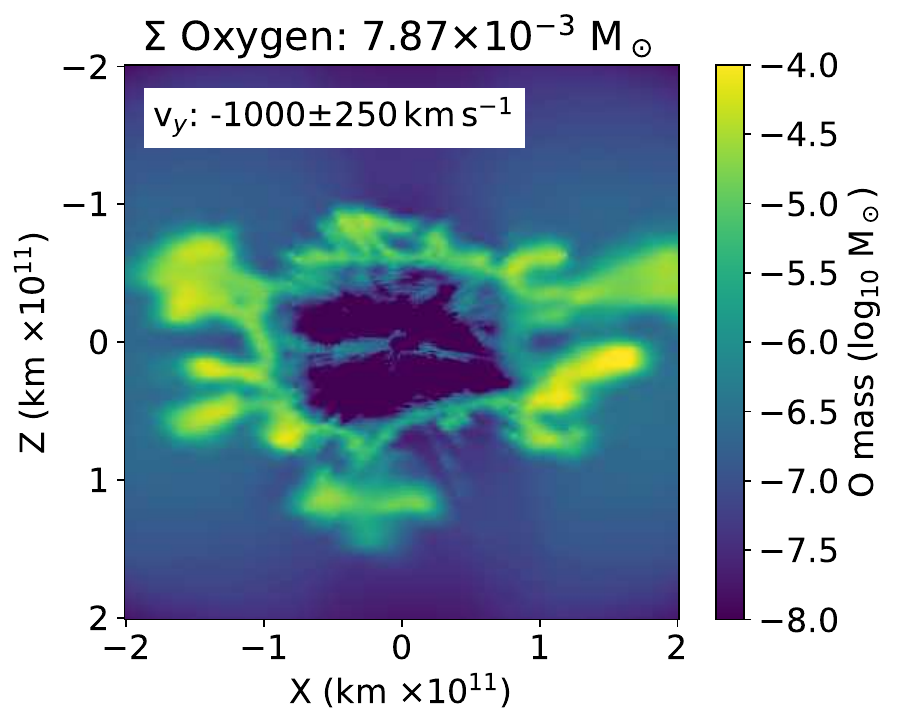}
    \includegraphics[width=.33\textwidth]{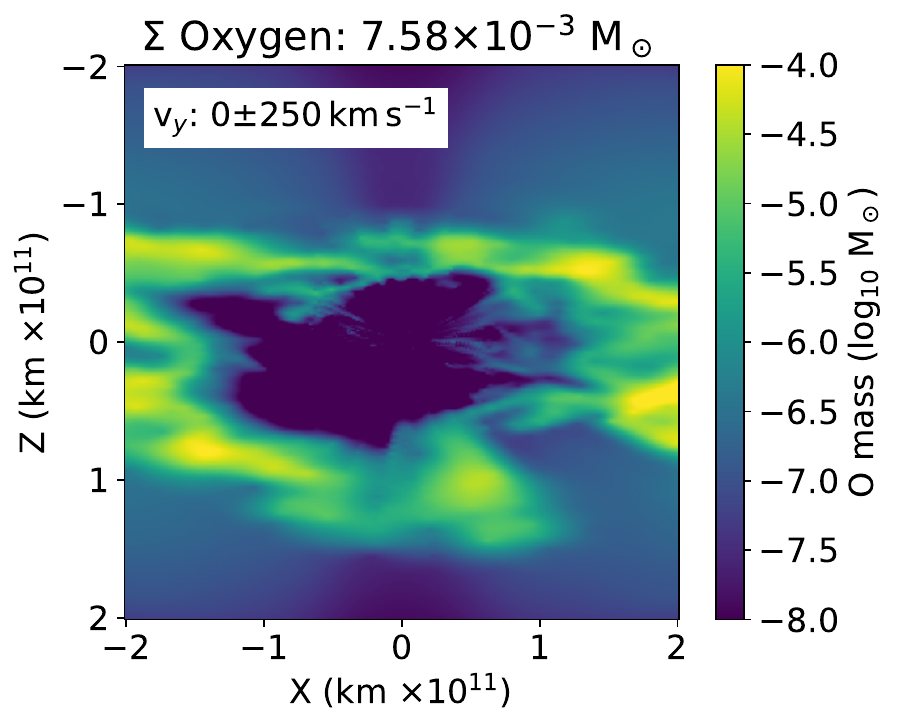}
    \includegraphics[width=.33\textwidth]{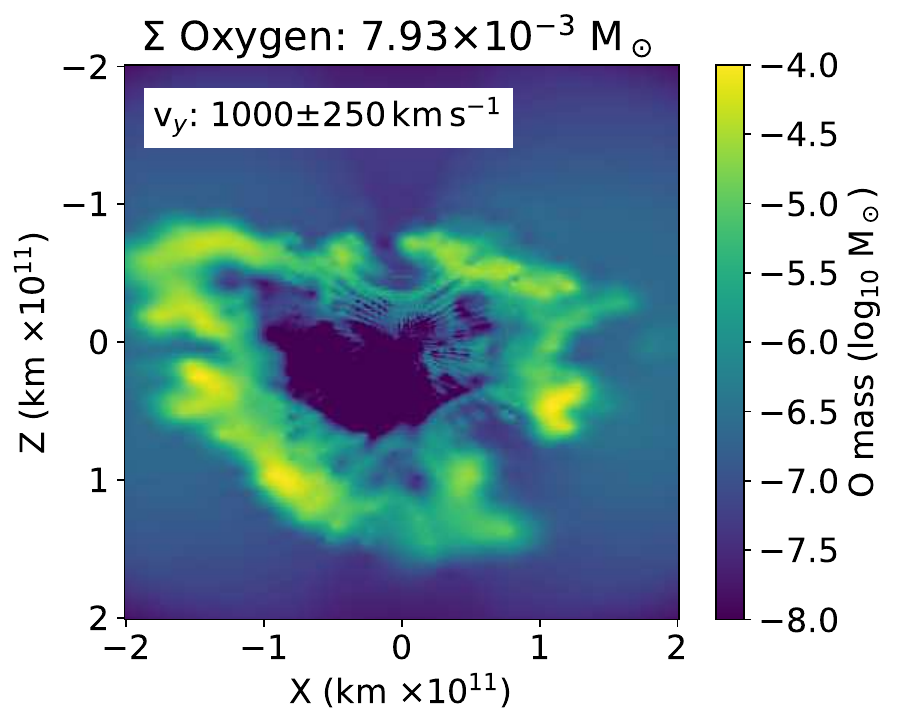}
    \includegraphics[width=.33\textwidth]{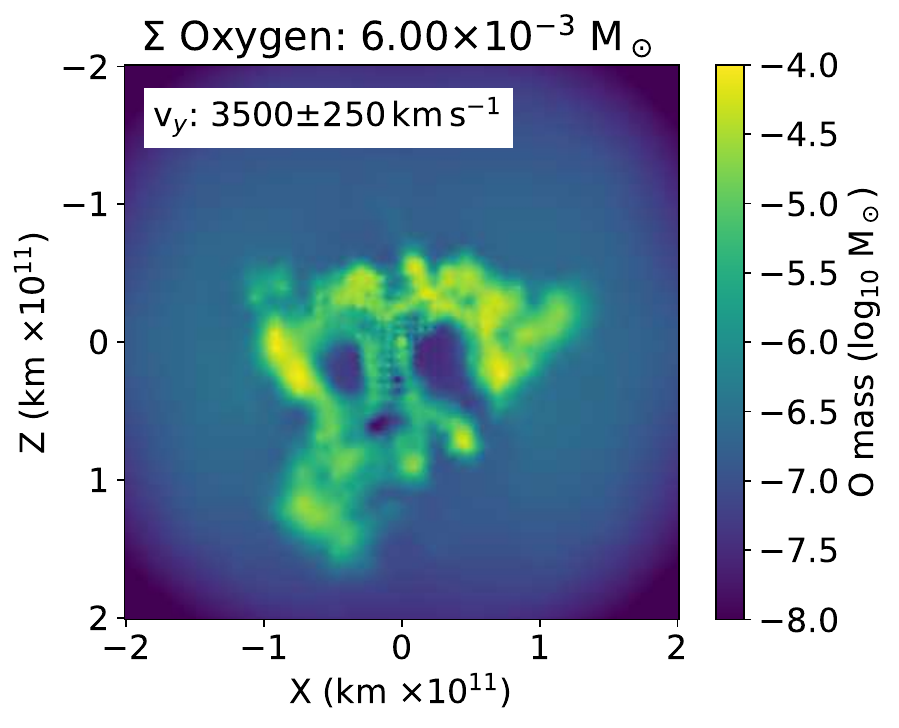}
    \includegraphics[width=.33\textwidth]{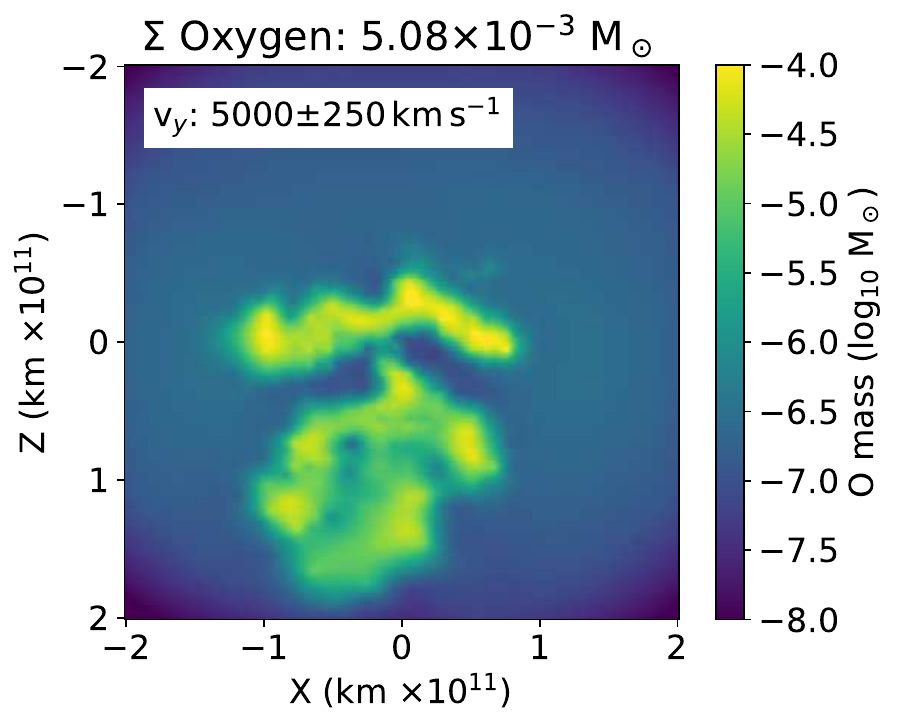}
    \includegraphics[width=.33\textwidth]{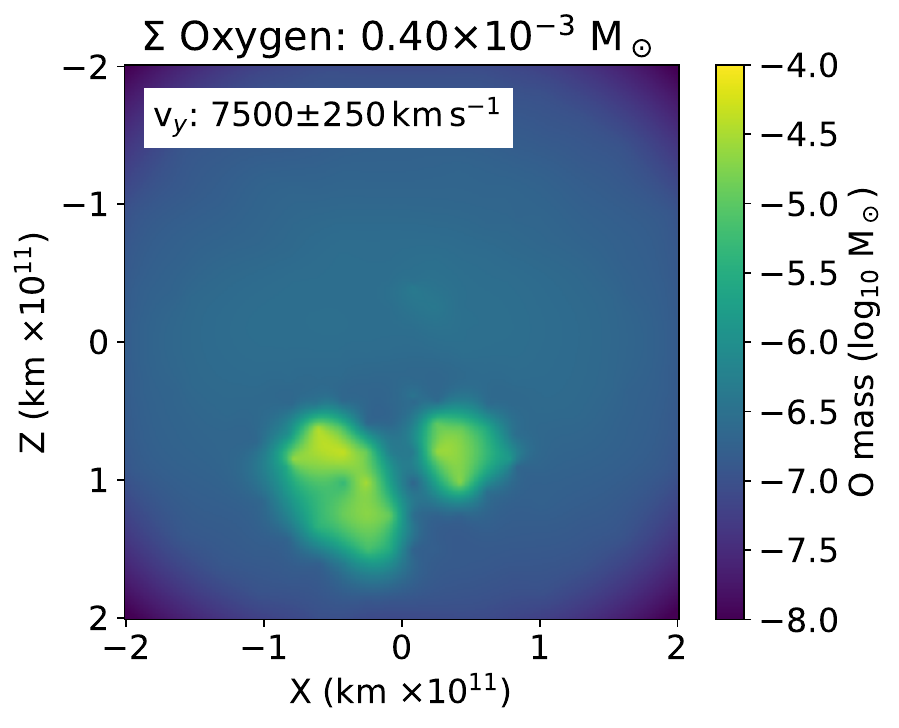}
    \caption{Nine different slice-throughs of the oxygen ejecta from the perspective of an observer looking straight down the y-axis of the model ($\theta=90^\circ\,\&\,\phi=0^\circ$, same as looking at Fig. \ref{fig:exploModel3D}). In each of the panels, the colour gradient goes from $10^{-8}\,$M$_\odot$ to $10^{-4}\,$M$_\odot$ and corresponds to the masses in the grid cells. The velocities ($v_y$) and total masses in the slices are marked with each panel. The X and Z bounds are set to $\sim0.04\,$c, as most mass in the system is limited to velocities ($v$) below $12,000\,$km$\,$s$^{-1}$. In the central, $v_y=0\,\text{km}\,\text{s}^{-1}$ panel, it can be seen that there is some material which has a very high $v_x$ which goes beyond this boundary. }
    \label{fig:OxygenSlices}
\end{figure*}
\begin{figure*}
    \includegraphics[width=.33\textwidth]{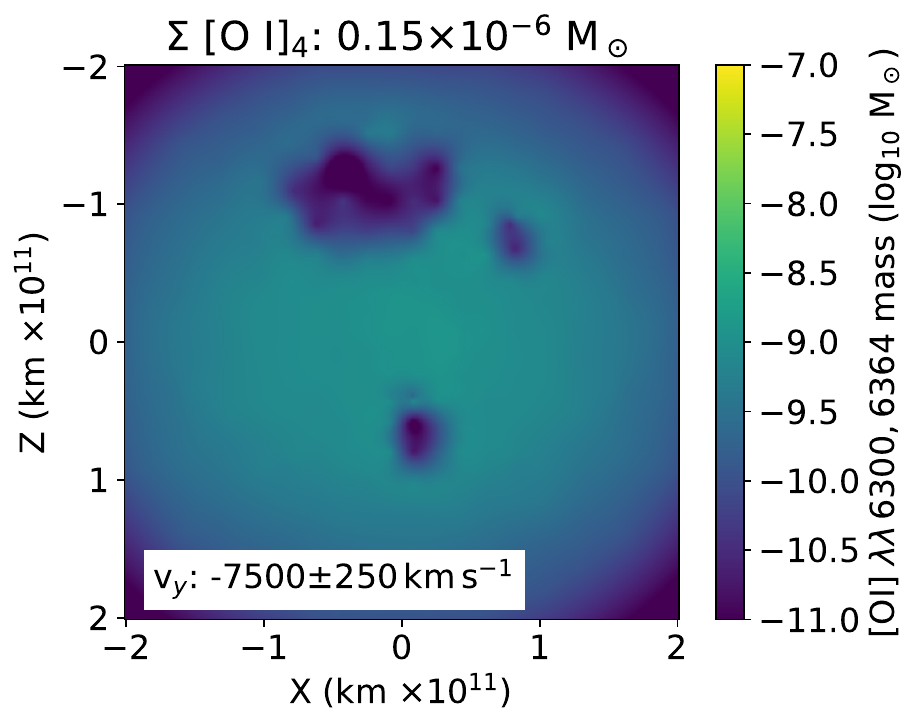}
    \includegraphics[width=.33\textwidth]{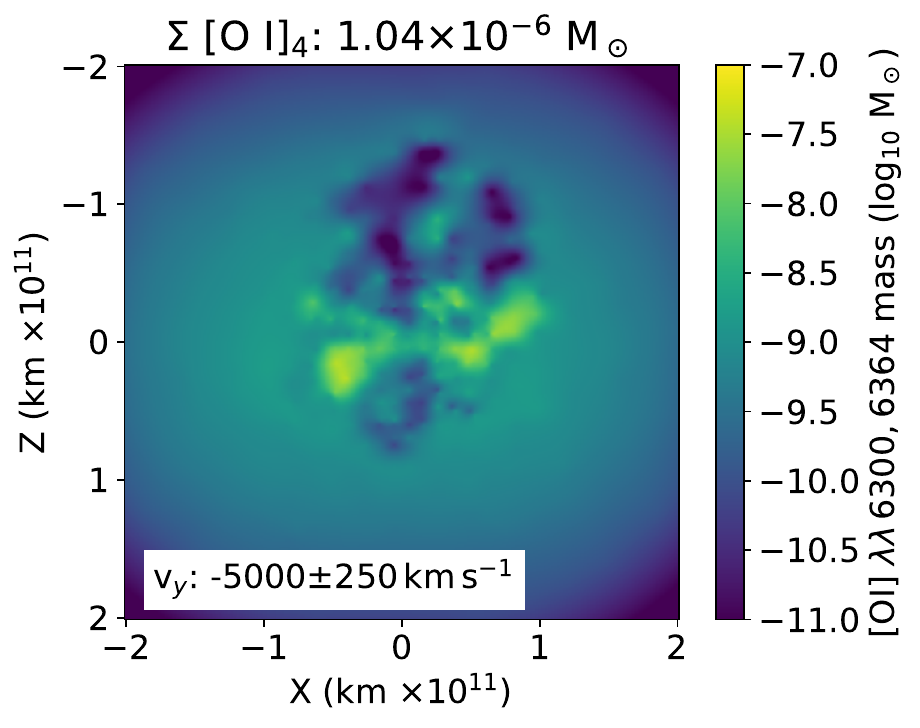}
    \includegraphics[width=.33\textwidth]{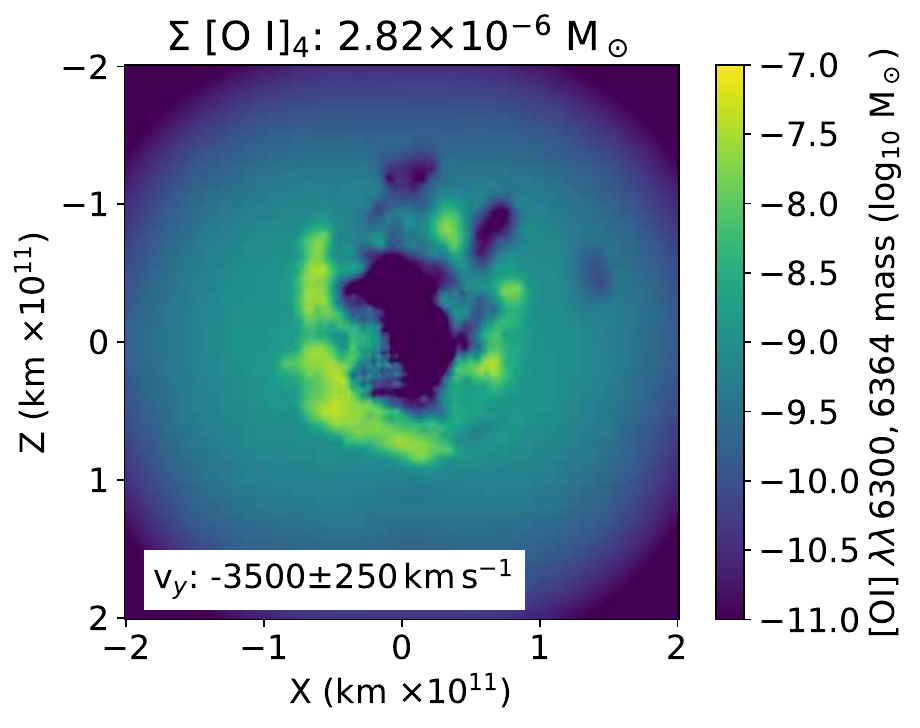}
    \includegraphics[width=.33\textwidth]{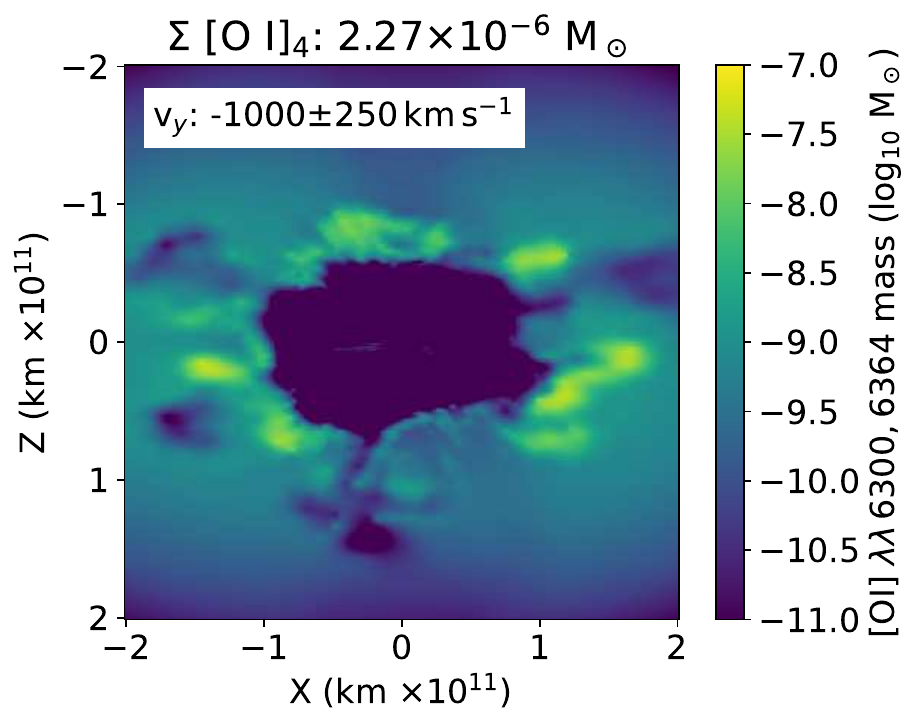}
    \includegraphics[width=.33\textwidth]{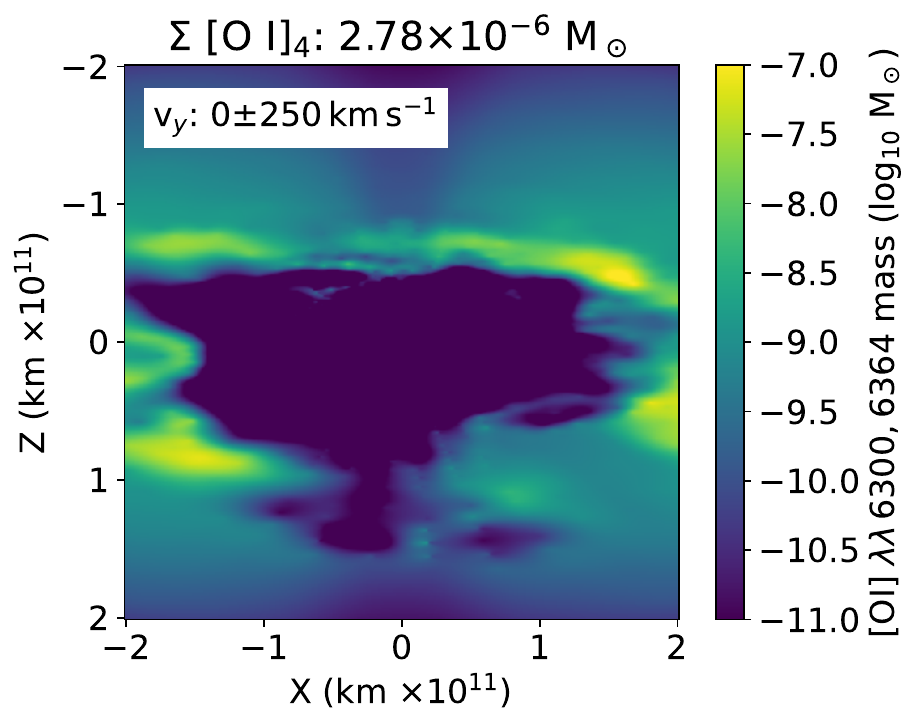}
    \includegraphics[width=.33\textwidth]{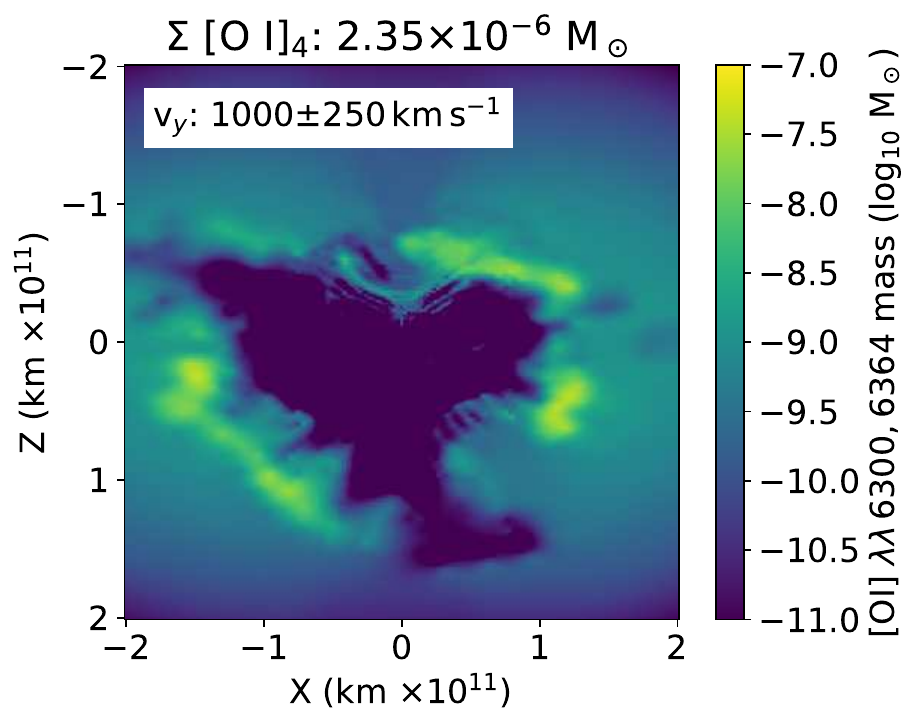}
    \includegraphics[width=.33\textwidth]{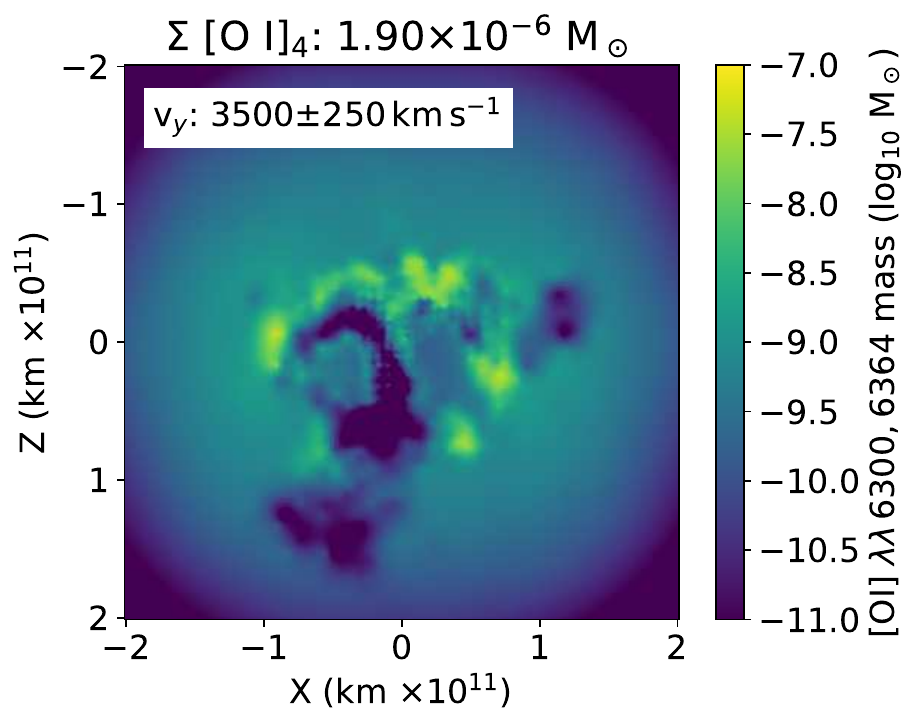}
    \includegraphics[width=.33\textwidth]{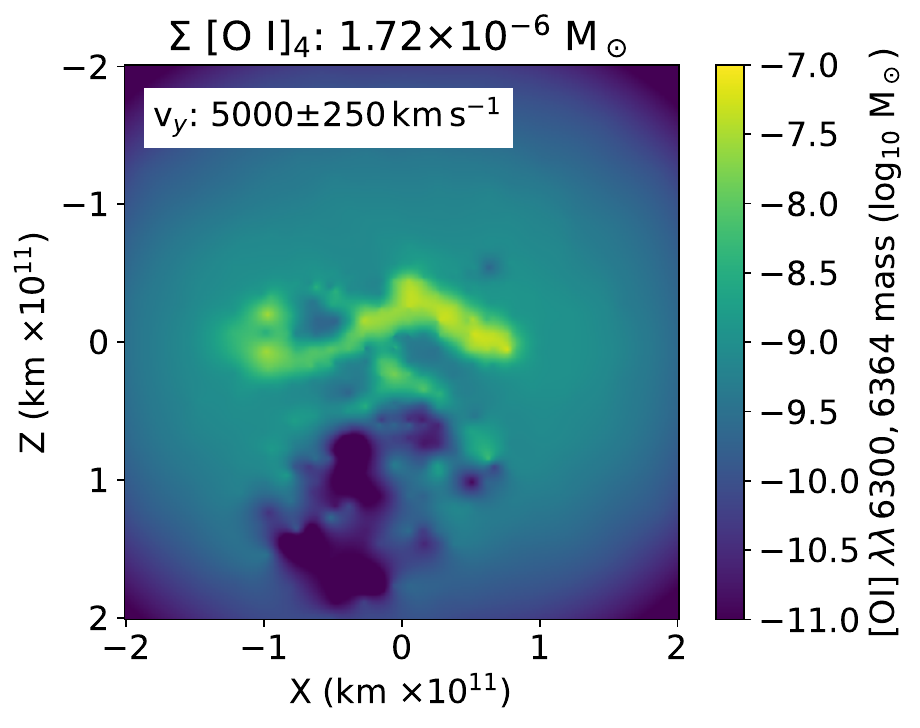}
    \includegraphics[width=.33\textwidth]{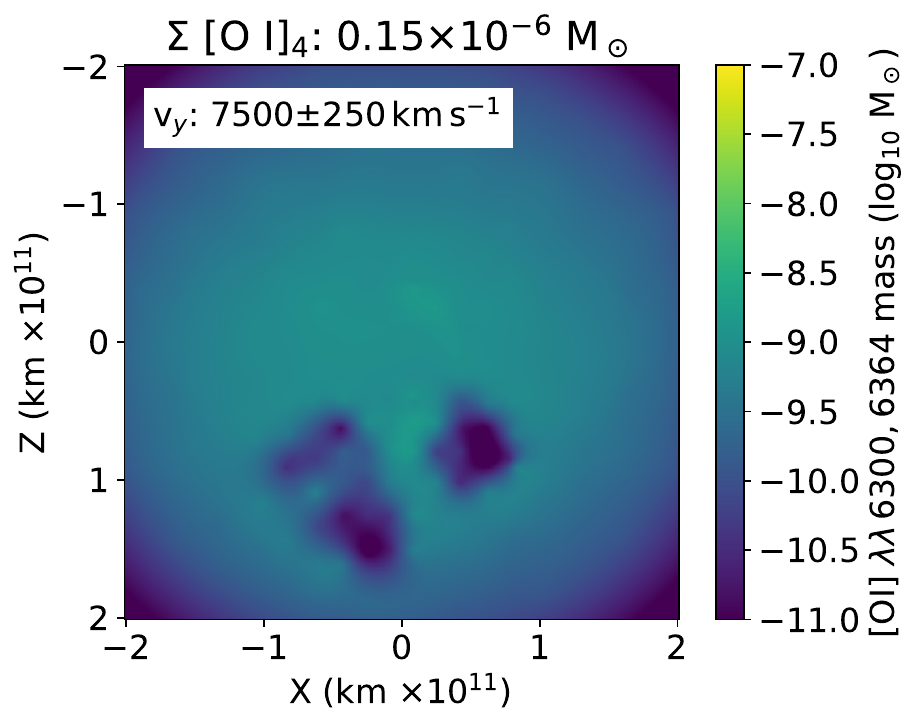}
    \caption{The same nine slices as in Figure \ref{fig:OxygenSlices}, but for the state giving [O I] $\lambda\lambda6300,\,6364$ emission. In each panel the colour gradient goes from $10^{-11}\,$M$_\odot$ to $10^{-7}\,$M$_\odot$ corresponding to the masses in the grid cells. The velocities ($v_y$) and total emitting masses in the slices are marked with each panel. The X and Z bounds are set to $\sim0.04\,$c, as most mass in the system is limited to velocities ($v$) below $12000\,$km$\,$s$^{-1}$.}
    \label{fig:ExcitationSlices}
\end{figure*}

In Figure \ref{fig:OxygenSlices} we display nine cuts through the ejecta along the X-Z plane, so on the same orientation as the 3D rendering in Figure \ref{fig:exploModel3D}. For each panel in these cuts, we took all the cells which have matching $v_y$ for that panel's velocity range, and added the mass of these cells at the right $x,z$ coordinates. An interpolator is then used for the purposes of smoothing the images.

The $y-$velocities are noted in each panel, starting at the slice the most 'out' of Figure \ref{fig:exploModel3D} and moving deeper with every panel from left to right, top to bottom. In each panel the mass located in each area is displayed on a log scale from $10^{-4}\,$M$_\odot$ to $10^{-8}\,$M$_\odot$, although the central cavities in the three middle panels go as low as $10^{-11}\,$M$_\odot$ (center) and $10^{-9}\,$M$_\odot$ (left$\&$right) in those cells. 

From these central panels it can be observed that material with a relatively low $v_y$ instead tends to have higher $v_x$ and/or $v_z$, as there is a large, empty cavity in the middle (as for tomography of a shell). Although those slices are the most symmetric ones, they are also clearly not spherical anymore, as the region with $z\leq-1\times10^{11}\,$km is practically empty compared to its positive-$z$ counterpart. The nine panels also indicate a somewhat elongated shape of the O-ejecta, as ejecta with negative $v_y$ is mostly found at negative $z$ while the positive $v_y$ material is predominantly on the positive $z$ side, although the masses at the extreme $v_y$ become very low.

The emission of the [O I] $\lambda\lambda 6300,\,6364$ doublet however comes only from the 4th excited state (the 1D state), so in Figure \ref{fig:ExcitationSlices} we display the same nine cuts through the ejecta as in Figure \ref{fig:OxygenSlices}, but only showing the oxygen which is in this excited state ([O I]$_4$). Several differences can be noted; in particular for the oxygen at high velocities, the [O I]$_4$ fraction is very small while the central cavities in the central panels also are much bigger for [O I]$_4$. This highlights two regions which hold significant oxygen mass but will contribute very little to the emission of the [O I] doublet.

Overall the mass in the emitting state is roughly a factor $2000-3000$ smaller than the total oxygen mass. Spatial variation of this factor will drive differences between the emergent line profile and the line profile one would obtain under the assumption that all oxygen emits equally. As can be seen in Figure \ref{fig:ExcitationSlices} this leads to more asymmetries compared to the full oxygen slices in Figure \ref{fig:OxygenSlices}, in particular for material with a high $|v_z|$ which contains very little emitting oxygen. This can be explained by the variation of gamma deposition and temperature throughout the ejecta, which governs the fraction of oxygen in the 4th state. Between Figures \ref{fig:OxygenSlices} and \ref{fig:ExcitationSlices} it also becomes clear that far from all oxygen is relevant to the emission lines, and thus that seeming mismatches between the NS-viewing angle $\Psi$ and redshits/blueshifts can appear. Such mismatches are displayed in Figure \ref{fig:lineprofiles_all}, where viewing angles which observe the NS coming towards them also observe the [O I] emission line as blueshifted. It might have been expected that the distribution of emitting [O I] would simply trace the bulk oxygen, but we find that this is not the case (see Figures \ref{fig:ExcitationSlices} and \ref{fig:OxygenSlices}).

\subsection{Momenta variations} \label{ssec:momenta_angles}
\begin{figure*}
    \includegraphics[width=\textwidth]{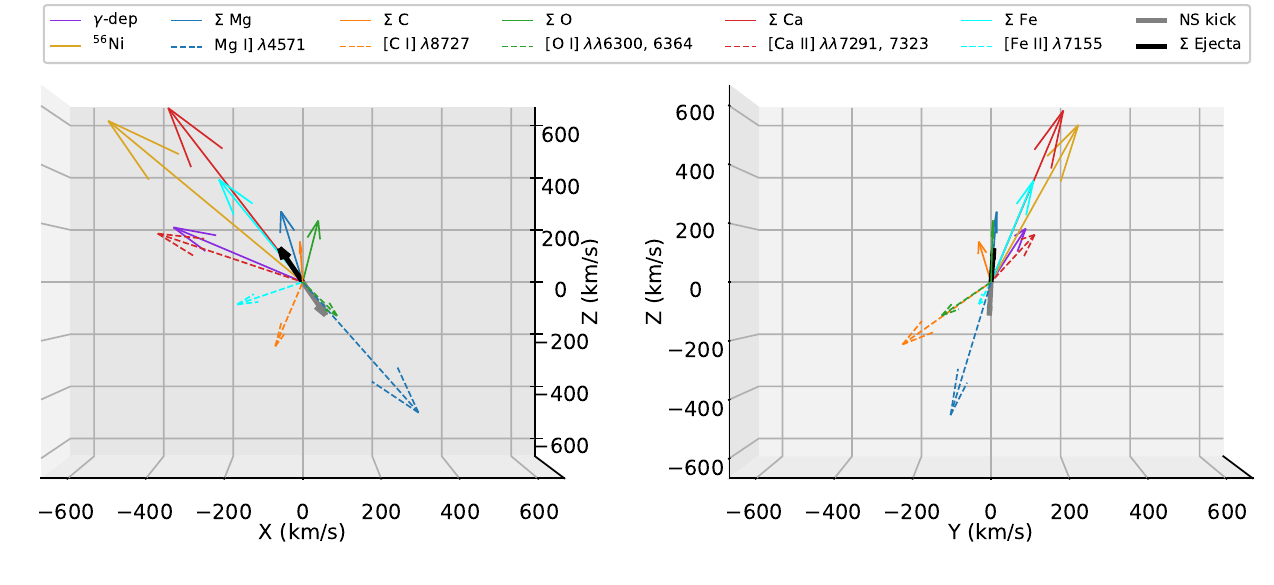}
    \caption{The momenta vectors for the NS (thick gray), total ejecta (thick black), $^{56}$Ni (solid gold), $\gamma$-deposition (solid purple), and Mg (blue), C (orange), O (green), Ca (red) and Fe (teal), with the solid lines corresponding to the momenta of each complete element, and the dashed lines to the excited states responsible for the marked features. The left shows the X/Z plane (at $v_y$ = 0) and the right the Y/Z plane (at $v_x=0$). The panels are at a $90^\circ$ rotation with the positive $x$-axis coming out of the paper.}
    \label{fig:momentavectors}
\end{figure*}
This clash of low $\Psi$ angles resulting in blueshifted emission lines is rather unexpected, as generally conservation of momentum leads to the idea that the neutron star and ejecta have to go (roughly) opposite ways. In Figure \ref{fig:momentavectors} we therefore showcase a series of momenta vectors to showcase that overall the ejected material is indeed moving in the opposite direction of the NS (solid black versus solid gray), and that for individual elements (solid lines) this also tends to be true, although in particular oxygen (green) is moving at a relatively different angle. It can also be seen that the $\gamma$-deposition vector (purple) is quite well-aligned with the $^{56}$Ni which is the radioactive power source, which is also to be expected.

The real surprise comes with the dashed lines, which correspond to the excited states responsible for the features shown here (blue for Mg I] $\lambda4571$, orange for [C I] $\lambda8727$, green for [O I] $\lambda\lambda6300,\,6364$, red for [Ca II] $\lambda\lambda7291,\,7323$). It can immediately be seen that indeed for Mg I] the emitting material is strongly co-moving with the NS, and that [C I] and [O I] are also roughly tracing the NS momentum and not the overall ejecta momentum. The outlier here is [Ca II], which might also explain why the $v_\text{shift}$, $v_\text{width}$ and $v_\text{skew}$ for [Ca II] are so different from the other three elements in Figures \ref{fig:lineprofiles_all} and \ref{fig:skewness_combi}. 

Additionally, the [O I] and [C I] vectors seem to be overlapping in the right hand plot while they not for the left hand plot, which can help explain why for some viewing angles their line profiles are good matches while for others they are quite poor, and why in particular these two elements might have good line profile matches.

Figures \ref{fig:OxygenSlices} and \ref{fig:ExcitationSlices} also show that across the ejecta, there are plenty of oxygen-rich areas which do not contribute to the [O I] $\lambda\lambda6300,\,6364$ emission at all, which helps explain the stark differences between the full vector and emitting vector in Figure \ref{fig:momentavectors}.

\citet{jerkstrand2020properties} looked at the $\gamma$-ray decay lines and iron group lines and the correlation between their centroid shifts and the NS kick vector, and found that these are also close to perfectly anti-aligned in their models. To compare against these results we also include Fe (teal) in Figure \ref{fig:momentavectors} and compare to [Fe II] $\lambda7155$ as emitting line only. We do find that this line is relatively anti-aligned to the NS kick compared to the other emission features bar [Ca II], although it is very weak in our spectrum (see Figure \ref{fig:HEC-33E_spectrum_per_el}).

\section{Observational Comparison} \label{sec:ObsComp}
\begin{figure}  
    \includegraphics[width=\linewidth]{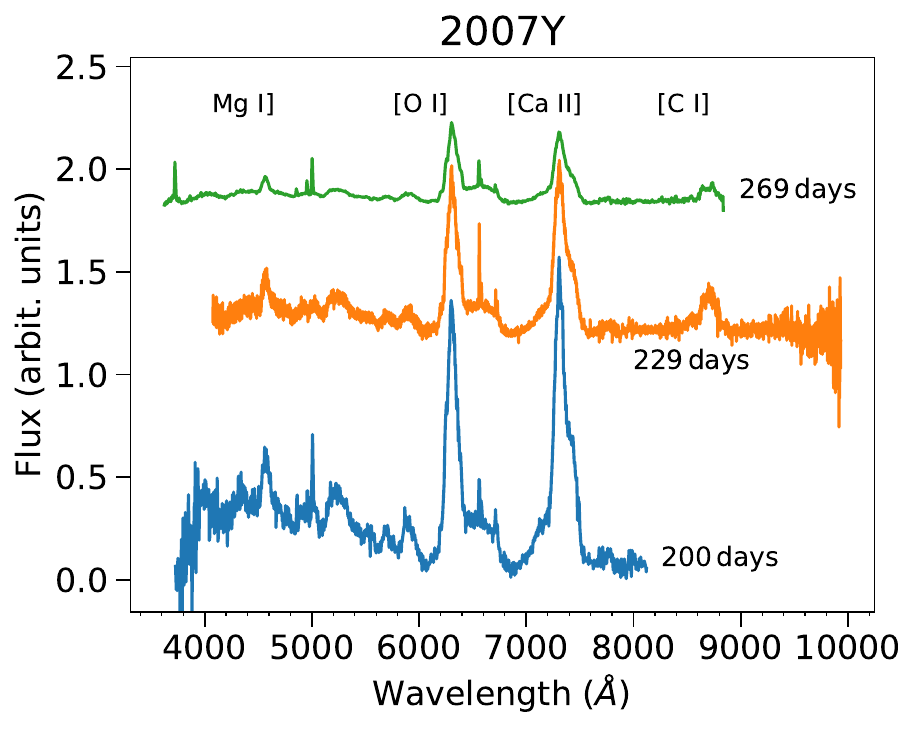}
    \caption{The observed nebular phase spectra of SN 2007Y, from \citet{stritzinger2009herich}.
    }
    \label{fig:SN2007Ystacked}
\end{figure}
\begin{table*}
	\centering
	\caption{Line centroid shifts ($v_\text{shift}$), line widths ($v_\text{width}$) and skewness values ($v_\text{skew}$) for SN 2007Y, for the three epochs of $200\,$days, $229\,$days and $269\,$days. The properties are determined by taking the flux present within $\pm5000\,\text{km}\,\text{s}^{-1}$ of the rest wavelength of each feature, corrected with a background noise subtraction. For the doublet features [O I] and [Ca II] the flux is taken from $-5000\,\text{km}\,\text{s}^{-1}$ from the blue side wavelength, up to the flux at $5000\,\text{km}\,\text{s}^{-1}$ from the red side, and the centroids are determined from the weighted average between the red and blue wavelengths ($6316\,\angstrom$ for [O I], and $7304\,\angstrom$ for [Ca II]). The spectra themselves come from \citet{stritzinger2009herich} and are shown in Figure \ref{fig:SN2007Ystacked} here. The measured shifts are mostly constant in time, although some variations are caused by a combination of noise and time evolution in the line or in contaminating lines.}
	\label{tab:2007Ycomparetable}
    \begin{tabular}{l|ccc ccc ccc} 
        \hline
        Feature & \multicolumn{3}{c}{$200\,$days} & \multicolumn{3}{c}{$229\,$days} & \multicolumn{3}{c}{$269\,$days} \\
          & $v_\text{shift}$ & $v_\text{width}$ & $v_\text{skew}$ & $v_\text{shift}$ & $v_\text{width}$ & $v_\text{skew}$ & $v_\text{shift}$ & $v_\text{width}$ & $v_\text{skew}$  \\
          & (km$\,$s$^{-1}$) & (km$\,$s$^{-1}$) & (km$\,$s$^{-1}$) & (km$\,$s$^{-1}$) & (km$\,$s$^{-1}$) & (km$\,$s$^{-1}$) & (km$\,$s$^{-1}$) & (km$\,$s$^{-1}$) & (km$\,$s$^{-1}$) \\ 
        \hline
		Mg I] $\lambda4571$ & -390 & 3998 & -1382 & -506 & 3670 & -1291 & -151 & 4296 & 1089 \\
		$[$O I] $\lambda\lambda6300,\,6364$ & -248 & 5215 & 1398 & -236 & 4862 & 861 & -171 & 5033 & 966 \\
		$[$Ca II] $\lambda\lambda7291,\,7323$ & 540 & 4513 & 1370 & 420 & 4196 & 1225 & 562 & 4411 & 1312 \\
		$[$C I] $\lambda8727$ & $-$ & $-$ & $-$ & -314 & 4625 & 1407 & -596 & 4571 & 966 \\
		\hline
	\end{tabular}
\end{table*}
We now turn to making some rough comparisons between our model line profiles and observations. As our model has an ejecta mass of $M_\text{ej}\approx1.3\,M_\odot$, which is relatively low even for a Type Ib SN (see e.g. \citealt{prentice2019investigating} who estimate typical $M_\text{ej}=2.8\pm1.5M_\odot$ for stripped envelope SNe, or the ejecta mass probabilities in \citealt{lyman2016bolometric} (Figure 10) for stripped envelope SNe), we are somewhat restricted in finding SNe with similar estimated ejecta masses which have good observational spectra past 200 days and which are clear of other powering mechanisms than radioactivity.

The best candidate we have found is SN 2007Y, which has nebular phase spectra taken at $200, 229$ and $269\,$days \citep{stritzinger2009herich} and which has low ejecta mass estimates while not being a potential magnetar, PWN or CSM interacting source. \citet{stritzinger2009herich} put the ejecta mass at $0.42\,M_\odot$, of which $0.2\,M_\odot$ is Oxygen, plus some unknown amount of $M_\text{ej,He}$ $-$ which is strikingly similar to the $0.42\,M_\odot$ of non-Helium ejecta we have, of which $0.16\,M_\odot$ is O. They furthermore also suggest that the progenitor star of this SN was a $3.3\,M_\odot$ He-core star, which is also the starting mass for our model. Other estimates for the ejecta mass of SN 2007Y are $1.4^{+1.3}_{-0.4}\,M_\odot$ \citep{lyman2016bolometric}, $1.9\,M_\odot$ \citep{taddia2018carnegie}, and $1.2\,M_\odot$ \citep[hypothesized to originate from a $3.4\,M_\odot$ He-core,][]{woosley2021model}, clearly putting it in the right range. A counterpoint against SN 2007Y is that the estimated energy is of order a few $\times10^{50}\,$erg, while our explosion model has $E_{kin}=1.05\times10^{51}\,$erg and thus might be too energetic. 

The three nebular phase spectra from SN 2007Y are shown in Figure \ref{fig:SN2007Ystacked}, corrected for the redshift of the host galaxy ($z=0.04657$). The regions of the four main emitting features from our explosion model (Mg I] $\lambda4571$, [O I] $\lambda\lambda6300,\,6364$, [Ca II] $\lambda\lambda7291,\,7323$ and [C I] $\lambda8727$) are also marked. The $200\,$day spectrum does not cover the [C I] feature and has an equivalent width resolution of $110\,\text{km}\,\text{s}^{-1}$ at $5000\,\angstrom$, while the $269\,$day spectrum marginally covers the [C I] feature and has a resolution of $99\,\text{km}\,\text{s}^{-1}$ at $5000\,\angstrom$. The $229\,$day spectrum is split into two halves (both orange in Figure \ref{fig:SN2007Ystacked}, they overlap at $\sim5500\,\angstrom$), with the first part covering the Mg I] feature with a resolution of $20\,\text{km}\,\text{s}^{-1}$, while the second part covers the other three features and has a resolution of $35\,\text{km}\,\text{s}^{-1}$, both if measured at $5000\,\angstrom$. In each of the spectra there is a sharp feature created by H$\alpha$ from the host galaxy, but it does not interfere with any of the features relevant to us. The broad feature on the red wing of the [O I] doublet is likely coming from [N II] $\lambda\lambda6548,\,6583$ \citep{jerkstrand2015late}, to which we cannot compare our HEC-33 model as N is not used in our nuclear network. 

For the spectra on display in Figure \ref{fig:SN2007Ystacked}, not all features have a proper baseline set, in particular for the $200\,$day spectrum this can obviously be seen for the Mg I] and [O I] features where the red and blue side of the features are not 'flat'. As such, when we want to determine the overall line centroid shifts and line widths for these features we first have to perform such a background subtraction ourselves, and for completeness sake we do this (separately) for every feature. For this background subtraction we take the average flux of a few bins just around $\pm5000\,\text{km}\,\text{s}^{-1}$ and draw a straight line between those points, which will serve as the baseline from which we measure the flux emitted by the feature. 

Once the proper flux regions have been chosen, we calculate the line centroid shifts, line widths (FWHM) and skewness the same as for our synthetic spectra, by using Equations \ref{eq:centroidEQ}, \ref{eq:widthEQ} and \ref{eq:skewEQ}. This way we are comparing the same properties as when we calculate the values for our own spectra, although now we have to keep in mind that we cannot filter out specific elemental contributions as we could before. 

The outcomes for each of the nebular phase epochs for the four features is shown in Table \ref{tab:2007Ycomparetable}, with the centroid shift being calculated from the rest wavelength for the singlet features Mg I] and [C I], and for the weighted average of these rest wavelengths for the doublet features (calculated as if under then optically thin limit, so $6316\,\angstrom$ for [O I] which has a 3:1 ratio, and $7304\,\angstrom$ for [Ca II] which has a 3:2 ratio due to the different statistical weights of the upper levels). For the [O I] feature care has to be taken, as there is some blending from a [N II] $\lambda\lambda6548,\,6583$ feature at the red side.

Comparing the line widths ($v_\text{width}$) from Table \ref{tab:2007Ycomparetable} clearly shows that these values are much lower than the ones inferred for our model, which had line widths of order $\sim6500\,\text{km}\,\text{s}^{-1}$ for Mg I] and [C I], $\sim7500\,\text{km}\,\text{s}^{-1}$ for [O I] and $\sim6800\,\text{km}\,\text{s}^{-1}$ for [Ca II]. The inferred widths for SN 2007Y might display a larger spread but are still significantly lower, ranging from $3670\,\text{km}\,\text{s}^{-1}$ for Mg I] at $229\,$days to $5215\,\text{km}\,\text{s}^{-1}$ for [O I] at $200\,$days. This large discrepancy might have its origin in the difference in explosion energy, which is estimated to be a factor 2 to 10 lower for SN 2007Y than for our explosion model. 

The line centroid ($v_\text{shift}$) estimates in Table \ref{tab:2007Ycomparetable} are perhaps a bit more extreme than found for most viewing angles for our model, but not exceptionally so, as almost all features at all epochs have $|\text{centroid}| < 500\,\text{km}\,\text{s}^{-1}$ $-$ with the Mg I] feature from $229\,$days being marginally outside this range, and the [C I] from $269\,$days being hard to determine correctly as it is so close to the edge of the spectrum; only for [Ca II] at $200$ and $269\,$days the centroids might be unexpectedly large. The values for each feature shift around a bit between the three epochs, with [O I] varying from $-248\,\text{km}\,\text{s}^{-1}$ to $-236\,\text{km}\,\text{s}^{-1}$ to $-171\,\text{km}\,\text{s}^{-1}$ between the three epochs, [Ca II] changes from $540\,\text{km}\,\text{s}^{-1}$ to $420\,\text{km}\,\text{s}^{-1}$ and finally $562\,\text{km}\,\text{s}^{-1}$, so for the doublet features the variation is not too drastic even if the total values are a bit high for [Ca II]. Meanwhile, for Mg I] the line is relatively weak and thus minor flux variations can have a strong impact on the shifts, which might be why the centroid shifts from $-390\,\text{km}\,\text{s}^{-1}$ to $-506\,\text{km}\,\text{s}^{-1}$ and finally $-151\,\text{km}\,\text{s}^{-1}$ for the three epochs. Lastly, for [C I] there is only one reliable estimate at the 229 day epoch, which gives $-314\,\text{km}\,\text{s}^{-1}$. What is truly noticeable about these estimates is that Mg I], [O I] and [C I] all have a relative blueshift for the line center, while [Ca II] instead has a redshifted line center. It is possible that this might be due to a contamination of the [Ni II] $\lambda7378$ feature which we can also find in our synthetic spectra in Figure \ref{fig:HEC-33E_spectrum_per_el}. On the other hand, this redshift/blueshift centroid dichotomy can partially also be observed for our own explosion model from Figure \ref{fig:lineprofiles_all}, where in particular for the red, orange and green dots ($\Psi\leqslant90^\circ$) in the Mg I], [O I] and [C I] profiles tend to favor centroid shifts on the blue side, while for [Ca II] those angles can predict a redshifted centroid or a blueshifted one, in particular for angles with $30^\circ\leqslant\Psi\leqslant90^\circ$. 

In Table \ref{tab:2007Ycomparetable} we also display the skewness ($v_\text{skew}$) of the line profiles, which is again fairly consistent for the doublet feature, as [O I] has values of $1398\,\text{km}\,\text{s}^{-1}$, $861\,\text{km}\,\text{s}^{-1}$ and $966\,\text{km}\,\text{s}^{-1}$ across the three epochs while [Ca II] has a skewness of $1370\,\text{km}\,\text{s}^{-1}$, $1225\,\text{km}\,\text{s}^{-1}$ and $1312\,\text{km}\,\text{s}^{-1}$. The skewness of the Mg I] feature is much more surprising, as the first two epochs find $-1382\,\text{km}\,\text{s}^{-1}$ and $-1291\,\text{km}\,\text{s}^{-1}$ respectively, while for the $269\,$day epoch we get $1089\,\text{km}\,\text{s}^{-1}$, so the overall skewness of the profile has flipped. Lastly, the [C I] feature at $229\,$days gives $1407\,\text{km}\,\text{s}^{-1}$ and $966\,\text{km}\,\text{s}^{-1}$ at the marginal $269\,$day spectrum. Excluding the first two Mg I] values, each profile in SN 2007Y has a positive skewness which is quite different from the profiles we calculated and show in Figure \ref{fig:skewness_combi}; we only find a few small regions where [O I] has a positive skewness and for those viewing angles at least one of the other elements is negatively skewed. For this observed spectrum such an opposite-skewness effect is not present as every feature (bar Mg I] in two epochs) is positively skewed. Mg I] might be so different as it is a relatively weak feature in the first two epochs with quite a bit of noise, which impacts our noise subtraction method and therefore bumps the skewness to negative values. 

Furthermore, in our model we calculate $v_\text{skew}$ by taking the line profile from $-10000\,\text{km}\,\text{s}^{-1}$ to $+10000\,\text{km}\,\text{s}^{-1}$, but for our observed spectra we use $\pm5000\,\text{km}\,\text{s}^{-1}$. This is because the lines are narrower for SN 2007Y than for our model and we want to avoid line blending, which is an issue with the observed spectra but not with the model spectra because for the model we can extract the flux from specific elements. This difference in the manner of calculating $v_\text{skew}$ might help explain why the values for the observed spectra are much lower than the extremes found in our synthetic ones.

The difference between the centroid shifts of the observed features might be an indication that there is some contamination in some of the observed profiles, from e.g. [Ni II] $\lambda7378$ in the [Ca II] profile, which does not appear for our synthetic spectra since we exclusively use the Ca-emission to determine our values. While it thus might be difficult to conclude that the centroid shifts calculated for [Ca II] are truly so different from the other elements, it should be noted that the line widths calculated do not appear to be very different from the other elements, which one might expect to happen if another element is blending into the line as that can have a broadening effect, and additionally the skewness values for the [Ca II] profile are also not very different from the other features.

\section{Discussion} \label{sec:Discussion}
When comparing our results to observations, ideally we compare against a SN which is as close to our explosion and ejecta parameters as possible, but as we only have one model available there are quite narrow restrictions on good-fitting observational SN. This will become easier in the future, when we can analyse more models with a much broader range of energies and ejecta masses. With those we will gain better understanding of what quirks of our model here might be model specific, or energy or mass specific, and which features and interpretations can be expected to be relevant for all Type-Ib SNe regardless of those parameter values.

In a similar vein, many nebular phase spectra of Type-Ib SNe can show oddities or unexpected similarities \citep[see e.g.][for previous studies on observed SNe and their line properties]{filippenko1989spectroscopic,maeda2008asphericity,modjaz2008double,taubenberger2009nebular} that we currently do not (fully) understand, which we might come to see in a new light once more different models have been tested and considered. Examples of such features might be the centroid shift differences between different elements $-$ is [Ca II] always going the opposite direction of the other main features (Mg I], [O I], [C I]), or only occasionally $-$ or is it an effect of [Ni II] $\lambda$7378 blending into the red wing? Does each of these features generally have a similar width in their profile, or is there some dependence in viewing angles together with ejecta mass or explosion energy? Are many of the small-scale wiggles that we observe in spectra due to noise in the detector, or are there actually microscopic details in the ejecta which causes those deviations that we can now uncover with our 3D setup? Many of these questions currently cannot be answered, but with \texttt{ExTraSS} and new models we will in the future gain the option to study such questions. 

Learning more about such questions will potentially help us gain an understanding on the late phases of stellar evolution $-$ and the agreement between future models and observations will be a key part in this. If such future models cannot accurately match the variations we see in observations, then somewhere along the way we are making a mistake or too gross of a simplification when calculating the models $-$ but that might be in the stellar evolution before the explosion, the details of the explosion itself, the first few seconds to minutes after the explosion or in the nebular phase modelling parts, which will make it difficult to trace back which component will have to be corrected. 

Regardless of what the future holds for making such comparisons with many models, with the model shown in this paper we can already see that there are many different kinds of line profiles possible for the ejecta of just one SN. This is potentially indicative of the fact that differences or similarities between observed line profiles for different SNe can perhaps simply be attributed to being a viewing angle effect, rather than some imprint of different kinds of SNe. 

One of the more interesting results we have found in this work is that emission lines can actually be aligned with the NS kick vector, which can have huge implications if emission lines are used as proxy for the momentum of the overall ejecta $-$ assumptions about which direction the NS might be going could be wrong simply because most of the ejecta is not emitting. The reason for this mismatch between the emission vectors and the elemental vectors is difficult to pin down, as the level populations of the emitting states depend on many factors, including the local temperature, free electron fractions and collisional rates. It is also possible that this appears in our model as the NS kick is relatively low ($135\,\text{km}\,\text{s}^{-1}$) and thus the ejecta might be relatively symmetric, and what are usually secondary components (e.g. mixing) suddenly become the primary components in setting the centroid shifts and line widths.

There are also several different improvements in the code that will be addressed in the future. Currently, \texttt{ExTraSS} does not do radiative transfer and we have already remarked in Section \ref{sec:Results} that this can potentially be the cause for the line strengths being hugely different from what is typically observed, as [O I] $\lambda\lambda6300,\,6364$ is nowhere close to being the strongest feature in our current setup, while that is typically the case in observed Type Ib nebular phase spectra. We also make the assumption of homologous expansion between the end-point of the hydrodynamical modeling from \texttt{Prometheus-HotB} to the timing epoch we use, which is several hundreds of days later. Work from \citet{gabler2021infancy} has shown that for H-rich SNe, dynamic evolution of the ejecta structure due to the $^{56}$Ni bubble expansion effect can take place over a timescale of weeks or longer. For a stripped-envelope SN as the one studied here, these inflation effects are expected to be smaller than for the H-rich SNe studied by \citet{gabler2021infancy}. This is because in a stripped envelope SN the metal layers, including the $^{56}$Ni, expands much faster than in a Type II SN. The higher velocities leads to a weaker $^{56}$Ni bubble effect in two ways. First, a lower ratio of the radioactive decay energy to the explosion kinetic energy. Second, a lower degree of gamma-ray trapping in the $^{56}$Ni clumps $-$ when clumps become transparent to the gamma-rays any pressure difference disappears and the inflation mechanism stops. However, some effect might still occur and would be interesting to study in a future work.

\section{Conclusions} \label{sec:Conclusion}
In this work we have presented a new, 3D nebular phase spectral synthesis code, \texttt{ExTraSS}, which can generate late-time NLTE spectra from multidimensional explosion models. The code is based on the 3D radiate transfer framework developed by \citet{jerkstrand2020properties}, incorporates some microphysics components from the 1D \texttt{SUMO} code \citep{jerkstrand201144ti,jerkstrand2012progenitor}, and combines these with several new treatments and modules to allow for full 3D modelling considering non-thermal physics and NLTE ionization and excitation. 

We apply the new code to a new 3D explosion model of a Type Ib supernova,  a $3.3\,$M$_\odot$ He-core progenitor (HEC-33) exploded with 1.05 Bethe, evolved to homology ($999\,$s) by the \texttt{Prometheus-HotB} code. We analyze the temperature and ionization in the ejecta, and their variation with position angle, over the period 150-300 days. In the regions rich in metals, for a given radial velocity  the temperature typically varies with position angle in the ejecta by about a factor 2, and electron fraction by about factor 1.5. For the He-rich outer layers the ejecta are closer to spherical symmetry and variations are smaller.

Our results show that there are large viewing angle deviations for the line profiles, with line profiles including Gaussian-like, horned, and strongly skewed ones, depending on viewing angle. We characterize the line profiles by three metrics; shift, width, and skewness, each of which diagnoses a particular aspect of the morphology. We use these to analyze the four strong lines of Mg I] $\lambda4571$, [O I] $\lambda\lambda6300,\,6364$, [Ca II] $\lambda\lambda7291,\,7323$ and [C I] $\lambda8727$. For the HEC-33 model, we obtain line widths between $6000-8000\,\text{km}\,\text{s}^{-1}$, centroid shifts of $-500$ to $+500\,\text{km}\,\text{s}^{-1}$, and skewnesses of $-3500$ to $3500\,\text{km}\,\text{s}^{-1}$, depending on viewing angle. These metrics define a domain in a 3-parameter space to which any observed SN lines can be compared.

We here compare the model to one particular supernova, SN 2007Y, which has been estimated to originate from a He-star with similar mass as our model. SN 2007Y shows nebular line shifts of a few hundred $\,\text{km}\,\text{s}^{-1}$ and line widths of $4000-5000\,\text{km}\,\text{s}^{-1}$. The observed line widths are a factor $\sim 1.5$ smaller than in the model, which seems in agreement with that the estimated explosion energy of the SN is 2-3 times smaller than our model. Correcting the line shift values by the same factor, the bulk asymmetries in SN 2007Y seem slightly larger than our model allows for. Skewness was more difficult to make direct comparisons for due to line blending.

Although there are similarities between the Mg I], [O I] and [C I] line profiles for many viewing angles, the four features investigated behave differently from each other, in particular [Ca II]. As each of these elements has a unique main burning process as its production channel, and therefore exists in particular regions in the SN, this is not surprising. However, our work is the first to quantity and analyse these similarities and differences based on 3D hydrodynamic models. 

A final intriguing outcome of our model is that the centroid shifts of the Mg I], [O I] and [C I] features are all in the same direction as the NS motion rather than opposite to it as the bulk element distributions. We believe that for this particular model, the spatial variation in physical NLTE conditions overwhelms the effect of overall element bulk shifts as these are quite small. This can be noted in particular in Figure \ref{fig:momentavectors}, where the emitting vectors are deviating strongly from the bulk vectors for the different elements. More extensive analysis of more models is needed to see how common this phenomenon is.


\section*{Acknowledgements}
The authors thank Quentin Pognan, Eliot Ayache, Conor Omand and Sofie Liljegren for their opinions and ideas during the development of this work, in particular QP for coining \texttt{ExTraSS}. The authors acknowledge support from the European Research Council (ERC) under the European Union’s Horizon 2020 Research and Innovation Programme (ERC Starting Grant No. [803189], PI: A. Jerkstrand). The computations were enabled by resources provided by the Swedish National Infrastructure for Computing (SNIC) at the PDC Center for High Performance Computing, KTH Royal Institute of Technology, partially funded by the Swedish Research Council through grant agreement no. 2018-05973. This work was supported by the German Research Foundation (DFG) through the Collaborative Research Centre ``Neutrinos and Dark Matter in Astro- and Particle Physics (NDM),'' Grant SFB-1258$\,$--$\,$283604770, and under Germany's Excellence Strategy through the Cluster of Excellence ORIGINS EXC-2094$\,$--$\,$390783311.

\textit{Software}: \texttt{Prometheus-HotB} \citep{fryxell1991instabilities,muller1991high,muller1991instability,janka1996neutrino,kifonidis2003nonspherical,kifonidis2006nonspherical,scheck2006multidimensional,arcones2007nucleosynthesis,wongwathanarat20133D,wongwathanarat2015three,wongwathanarat2017production,ertl2016two}, \texttt{NumPy} and \texttt{SciPy} \citep{numpyscipy}, \texttt{Matplotlib} \citep{hunter2007matplotlib}, \texttt{Pandas} \citep{pandas_2023_feb}.

\section*{Data Availability}

The Data underlying this article will be shared on reasonable request to the corresponding author.



\bibliographystyle{mnras}
\bibliography{NLTE} 




\appendix

\section{Ejecta details $\&$ homology}  \label{app:ejecta}
In this Appendix we will show some more details on the explosion model 'HEC-33' used as the basis for our nebular phase NLTE spectra, and we will also showcase some extra calculations to validate the homology assumption we make for the fast-forwarding from \texttt{P-HotB} to \texttt{ExTraSS}. 

\subsection{Progenitor details} \label{app:progenitor}
\begin{figure}
    \centering
    \includegraphics[width=\linewidth]{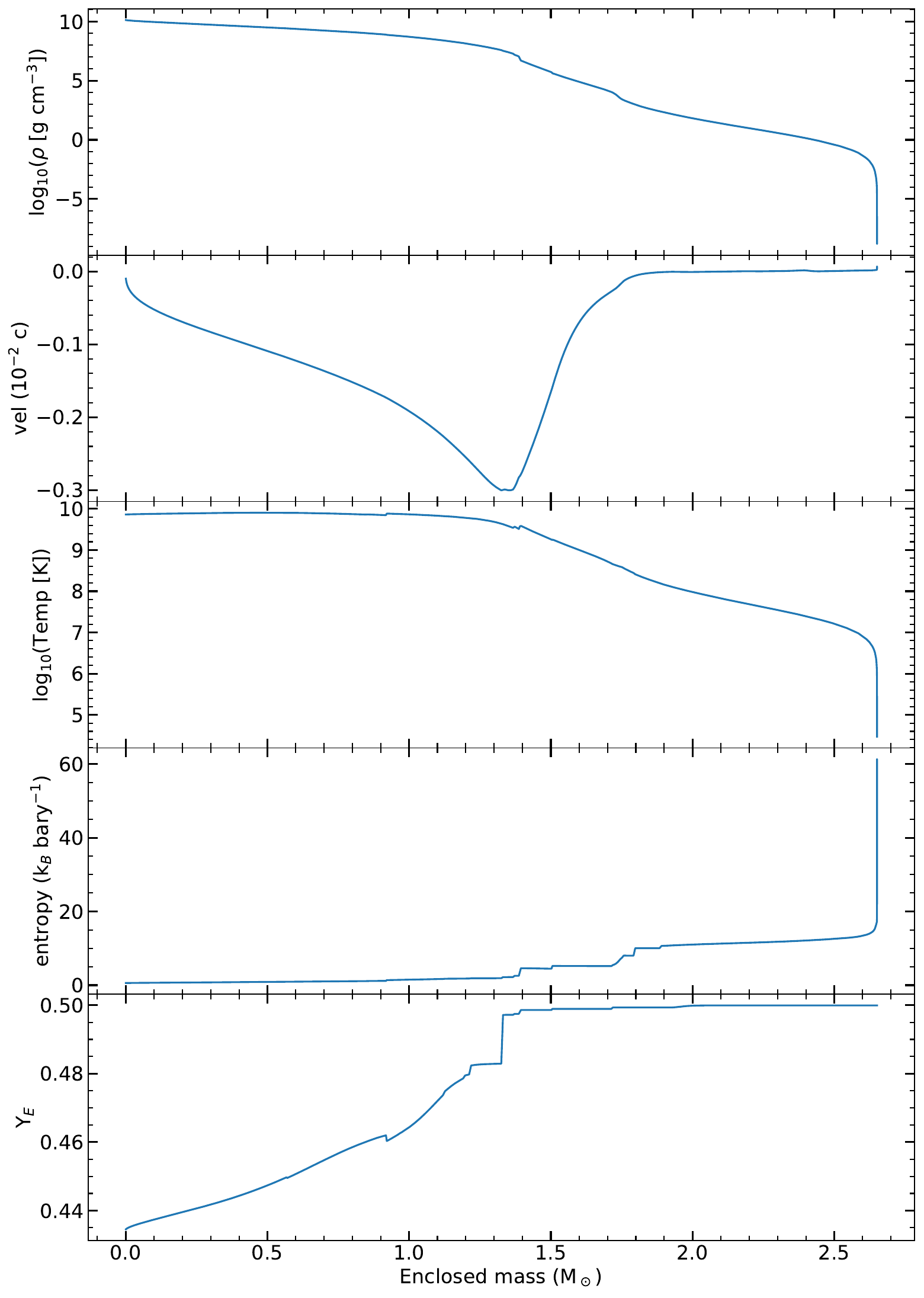}
    \caption{Radial profiles along the mass coordinate for density (top), velocity (2nd row), temperature (middle), entropy (fourth) and electron fraction ($Y_E$, bottom) for the progenitor of HEC-33 at the onset of iron-core collapse (as visible from the negative velocities signalling the beginning of the infall). The density and temperature profiles are on logarithmic scales.}
    \label{fig:prog_data_details}
\end{figure}
\begin{figure}
    \centering
    \includegraphics[width=\linewidth]{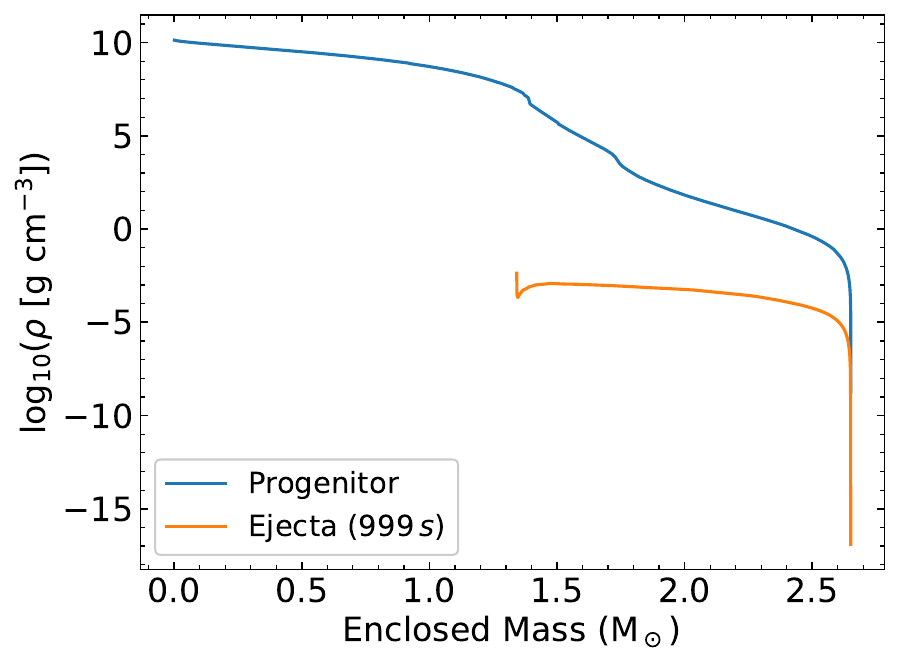}
    \caption{A comparison of the density profiles of the progenitor star (blue) before the explosion, and the ejecta (orange) at $t=999\,$s against their enclosed mass coordinate. The mass of the neutron star which is formed in the explosion is considered the starting point for the ejecta mass, rather than $M=0$.}
    \label{fig:density_both}
\end{figure}
The progenitor structure is a critical detail in this work as it serves for the basis of the explosion model, whose ejecta are used to fast-forward into the nebular phase and then obtain NLTE spectra with \texttt{ExTraSS}. In Figure \ref{fig:prog_data_details} we show some additional interior physical details, namely the density profile, radial velocity, temperature, entropy and electron fraction at the time of explosion, against the radial mass coordinate. Additionally, in Figure \ref{fig:density_both} the density profiles of the progenitor star and the ejecta (at $t=999\,$s) are shown together to highlight how much more diffuse the ejecta have become after explosion.

\subsection{Homology validation $\&$ fallback} \label{app:homology_confirm}
As mentioned in Section \ref{ssec:ejectamodel}, a critical point for our study is the hand-off between the hydrodynamic modelling in the explosion model with \texttt{P-HotB} and our new nebular phase NLTE code \texttt{ExTraSS}. In \texttt{ExTraSS} we assume homologeous expansion, i.e. everything in the ejecta is moving only in the radial direction with the velocity set as the free-coasting velocity $v_{fc}=r/t$, where $r$ is the distance from the centre and $t$ the time since the explosion. While it is possible to run the \texttt{P-HotB} code until later times (e.g. half a day post-explosion) it is also expensive to do so, and numerical diffusion by the fluid moving across the Eulerian grid may become more severe. If too early a snapshot is used, however, there are still significant non-radial motions present on the grid, which makes the assumption of free-coasting velocity worse to apply. 

We can directly test to what extent the velocity field in a given model is well described by a $v_{fc}$ law. The first check is that the angular velocities $v_{ang}(r)=\sqrt{v_\theta^2 + v_\phi^2}$ are much smaller than the radial velocities $v(r)$. The second is that the $v(r)$ distribution (for different angles) has only small variations at every $r$. If those tests are passed, a final comparison between $v(r)$ and $v_{fc}(r)$ can be made. For this we apply the following:
\begin{equation}
    \chi(r) = \dfrac{|v(r) - v_{fc}(r)|}{v_{fc}(r)},
    \label{eq:chi_mass}
\end{equation}
where $v(r)$ is the average radial velocity for the ejecta at every radius and $v_{fc}(r)$ is as given above. Thus from Equation \ref{eq:chi_mass} we obtain a value of $\chi$ for every radius in the model, upon which we apply a weighting for the amount of mass in each radius ($M(r)$) in the model to obtain a final value $\bar\chi$:
\begin{equation}
    \bar\chi = \dfrac{\Sigma \,[\chi(r) \times M(r)\,]}{\Sigma M(r)}.
    \label{eq:chibar_final}
\end{equation}
Attaining a low value for $\bar\chi$ without extending the duration of the hydro modelling unnecessarily is the main goal, as simply aiming for the lowest possible value of $\bar\chi$ would result in running \texttt{P-HotB} for as long as possible. Taking the mass-weighted average of $\chi(r)$ is important as typically for both low and high $r$ the radial velocity $v(r)$ has a bigger offset from $v_{fc}(r)$, but there is generally also relatively little mass found at those radii. 
\begin{figure}
    \includegraphics[width=\columnwidth]{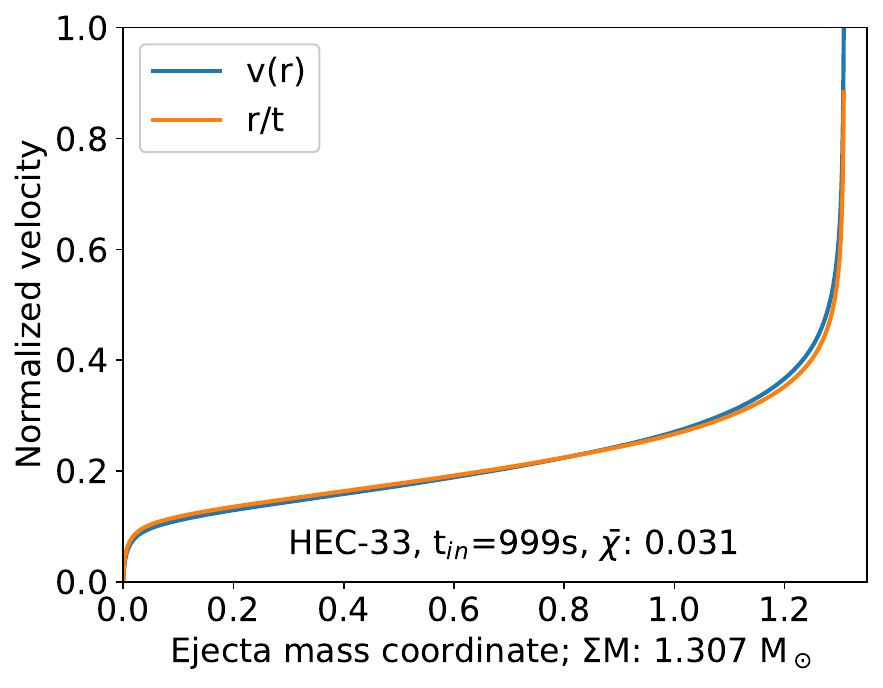}
    \caption{Normalized velocity curves of the angular mean of the velocity versus radius, $v(r)$ (as used in Equation \ref{eq:chi_mass}), and the quantity $v_{fc}(r) \equiv r/t$, both versus mass coordinate. The velocity curves are plotted to a cut-off point of $v(r)\leqslant36,000$ km s$^{-1}$. The value of $\bar\chi$, the level of agreement between the curves, is calculated using the mass-weighted average of Equation \ref{eq:chibar_final}.}
    \label{fig:HEC33_chibar_timing999}
\end{figure} 
In Figure \ref{fig:HEC33_chibar_timing999} we show the $v(r)$ and $v_{fc}(r)$ curves for the $3.3\,$M$_\odot$ He-core explosion model at $999\,$s post explosion, together with $\bar\chi$. Both the velocity curves are divided by $.121\,c$ as the radii past that velocity have not yet been passed by the expanding shock wave and thus do not contain any SN ejecta. It can clearly be seen that the areas with the worst matching of $v(r)$ and $v_{fc}(r)$ are also carrying the least mass (as the slopes are steepest there) and thus we obtain a reasonably low value of $\bar\chi=0.031$. The homology deviation of 3\% is small enough to warrant the use of the $999\,$s post explosion data from HEC-33 as input for \texttt{ExTraSS}.

\begin{figure}
    \centering
    \includegraphics[width=\linewidth]{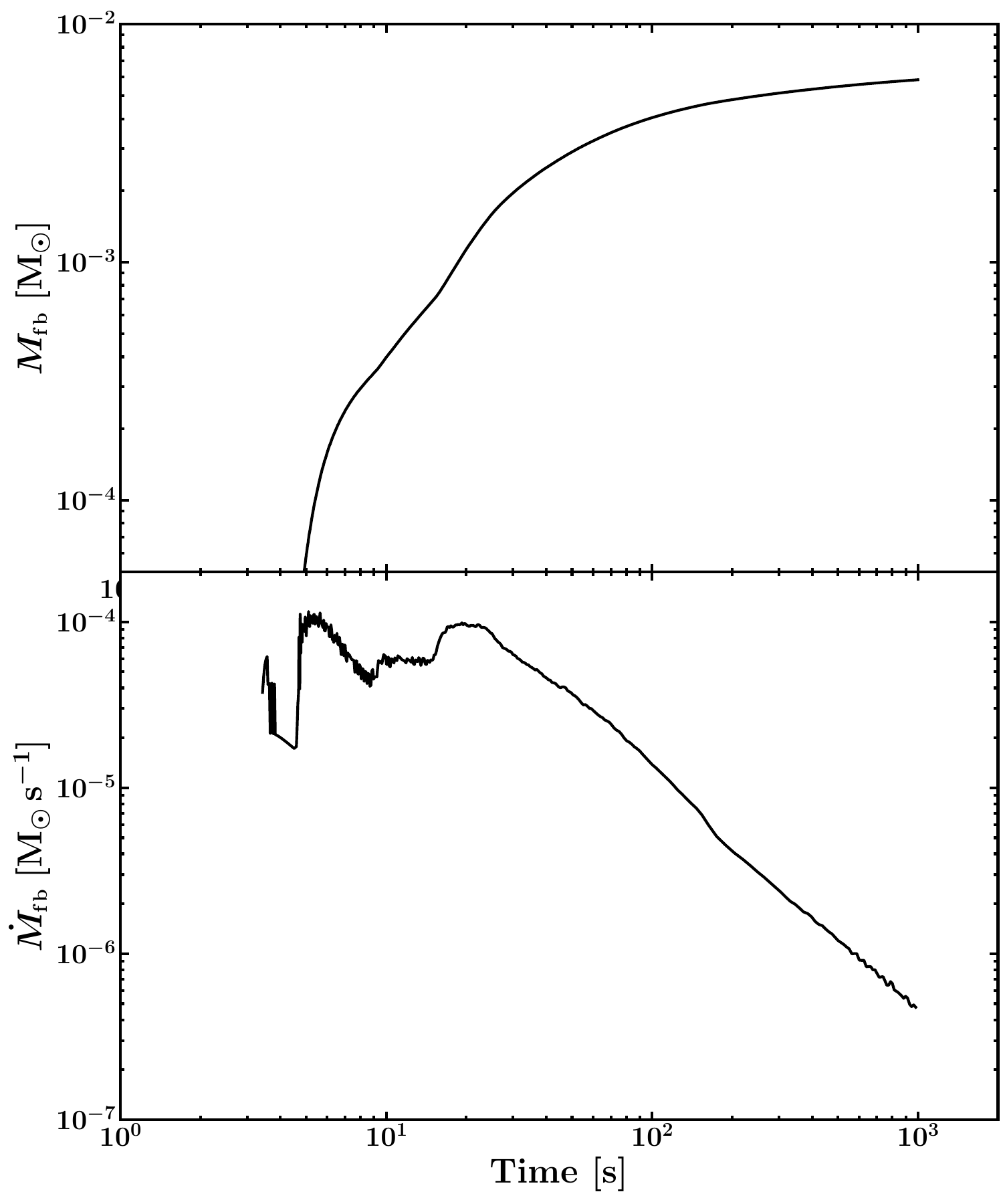}
    \caption{Cumulative fallback mass $M_\text{fb} (M_\odot)$ (top) and time-dependent fallback mass $\dot M_\text{fb} (M_\odot\,\text{s}^{-1})$ (bottom), against the time since explosion. After the first minute, $\dot M_\text{fb}$ has dropped well below $10^{-5}\,M_\odot\,\text{s}^{-1}$ and keeps steadily dropping until the end of the model at $t=999\,$s. The overall $M_\text{fb}$ stagnates at a few $10^{-3}\,M_\odot$.}
    \label{fig:fallback}
\end{figure}
Section \ref{ssec:ejectamodel} also mentioned that material which leaves the grid through the inner boundary during the \texttt{P-HotB} modelling is assumed to accrete back onto the central object as fallback. In Figure \ref{fig:fallback} we show both the time-dependent evolution of this fallback mass as well as the cumulative fallback for the HEC-33 model, up to $t=999\,$s. 

As can be seen, the overall fallback mass $M_\text{fb}$ remains well below $0.01\,M_\odot$ even at the end of the simulation and appears to be approaching an asymptotic limit. The steady decline of $\dot M_\text{fb}$ also indicates that almost all of the material which is ejected at $t=999\,$s will remain so and not fall back at a later time. The nearing of an asymptotic limit for $M_\text{fb}$ as well as the strong decline of $\dot M_\text{fb}$ are additional hints that the innermost ejecta has achieved a stable velocity and will continue moving outward, further reinforcing the idea that at $t=999\,$s the ejecta can be fast-forwarded under the free-coasting approximation.


\bsp	
\label{lastpage}
\end{document}